# Photometric activity cycles in fast-rotating stars: Revisiting the reality of stellar activity cycle branches

Deepak Chahal[1,2]★, Devika Kamath[1,2], Richard de Grijs[1,2,3], Benjamin T. Montet[4,5] and Xiaodian Chen[6]
[1]*School of Mathematical and Physical Sciences, Macquarie University, Balaclava Road, Sydney, NSW 2109, Australia*
[2]*Astrophysics and Space Technologies Research Centre, Macquarie University, Balaclava Road, Sydney, NSW 2109, Australia*
[3]*International Space Science Institute–Beijing, 1 Nanertiao, Zhongguancun, Hai Dian District, Beijing 100190, China*
[4]*School of Physics, University of New South Wales, Sydney, NSW 2052, Australia*
[5]*UNSW Data Science Hub, University of New South Wales, Sydney, NSW 2052, Australia*
[6]*CAS Key Laboratory of Optical Astronomy, National Astronomical Observatories, Chinese Academy of Sciences, Beijing 100101, China*

**ABSTRACT**
We aim to detect activity cycles in young main-sequence stars, analogous to the 11-year solar cycle, using combined photometric survey data. This research will enhance our understanding of how cycle periods relate to rotation rates in fast-rotating stars. We measured activity cycles for 138 G–K-type main sequence stars using combined time-series photometry spanning ∼14 years. The first set of 70 stars used data from *Kepler* Full Frame Images (FFIs)–ASAS-SN–ZTF, and the second set of 68 stars used data from *Kepler*-FFIs–ZTF. Additionally, we measured the activity cycles for 25 RS CVn candidates. For our sample, we analyzed the correlation or anti-correlation between flux variations and photospheric activity which arises due to presence of faculae or starspots. We identified fast-rotating K-type stars that are faculae-dominated by tracking spot/faculae evolution in *Kepler* RMS data. Our findings reveal that fast-rotating G–K-type stars shows no strong correlation between cycle length and rotation period. Previous studies have identified active and inactive branches in the cycle–rotation diagram. However, we find that G–K-type stars do not show a clear trend aligning with the active branch, with 34 per cent stars falling within the intermediate region between the two branches, where our Sun resides. Our results highlight that the proposed distinction between the two branches may not be as definitive as previously thought, particularly regarding the active branch. Furthermore, we also detected 23 per cent of young Sun-like stars in the intermediate region, where our Sun is located, implying that our Sun may not be unique in this regard.

## 1 INTRODUCTION

Stellar activity cycles, similar to the 11-year solar cycle, are (quasi-)periodic variations in the intrinsic brightness of stars driven by changing magnetic fields. Such stellar cycles are believed to be a result of dynamo processes. However, understanding the stellar dynamo, a process responsible for amplifying and sustaining stellar magnetic fields, remains a major challenge. Pioneering work by Wilson (1968, 1978) and others (Noyes et al. 1984; Baliunas et al. 1995; Brandenburg et al. 1998; Saar & Brandenburg 1999) has revealed chromospheric activity cycles in many stars, including the Sun. The solar cycle can be explained by an $\alpha\Omega$ dynamo (Baliunas et al. 1995; Dikpati & Gilman 2006; Charbonneau 2010), where latitudinal differential rotation transforms a poloidal field into a toroidal one (known as the '$\Omega$ effect'), and cyclonic convection (Parker 1955) or meridional circulation transporting flux to poles (Babcock 1961; Leighton 1964) then restores the poloidal field but with reversed polarity (known as the '$\alpha$ effect').

Understanding stellar cycles for different stellar masses and their connection to a range of fundamental stellar parameters (rotation, age, metallicity, etc.) is crucial for constraining stellar dynamo physics. These activity cycles are observed over long-term monitoring campaigns of activity indices, including chromospheric variability (traced by Ca II H and K; Baliunas et al. 1995; Oláh et al. 2016; Boro Saikia et al. 2018; Olspert et al. 2018; Isaacson et al. 2024) or coronal (X-ray) variability (Coffaro et al. 2020). However, characterising stellar cycles using spectroscopic data is challenging because of the requirement for long-term monitoring. Meanwhile, our Sun also changes its brightness by ∼0.1 per cent in the optical (Fröhlich & Lean 2004) during the solar activity cycle. Decades-long spectroscopic and photometric time-series observations have revealed stellar cycles similar to the 11-year solar cycle. Several recent studies have used All Sky Automatic Survey for SuperNovae (ASAS-SN) (e.g., Suárez Mascareño et al. 2016; Distefano et al. 2017; Irving et al. 2023) and *Kepler* (e.g., Montet et al. 2017; Reinhold et al. 2017) photometric data to characterise stellar activity cycles. Photometric variations associated with activity cycles have been detected even in fainter stars, such as in fully convective M dwarfs (Suárez Mascareño et al. 2016; Irving et al. 2023).

Observations of stellar cycles for stars at different evolutionary stages provide a valuable opportunity to study stellar dynamo action as a function of stellar age. Older stars, characterised by slower rotation rates ($P_{\rm rot} \geq 20$–25 days) exhibit longer activity cycles ($P_{\rm cyc} \geq 5$–7 years) with lower cycle amplitudes (Noyes et al.

★ E-mail: deepakchahal294@gmail.com

© 2015 The Authors

1984; Baliunas et al. 1995). Dynamo processes are expected to be different in fast- and slowly rotating stars (Kitchatinov & Rüdiger 1999). However, more recent studies (Lehtinen et al. 2016; Suárez Mascareño et al. 2016) have demonstrated the presence of similar cycle periods in both fast- and slowly rotating stars. In some active stars, multiple cycles have also been detected (Berdyugina & Tuominen 1998; Berdyugina & Järvinen 2005). Brandenburg et al. (2017) discovered shorter cycles ($P_{cyc} \sim 1$–3 years) coexisting with longer cycle periods in fast-rotating stars, indicating the simultaneous operation of two underlying dynamo processes. Moreover, a recent study by do Nascimento et al. (2023) revealed that a Sun-like star, 18 Scorpii (18 Sco), exhibits Sun-like twin cycles of $\sim 6.9$ and 15 years, similar to the solar Schwabe (11-year sunspot) and Hale (22-year polarity reversal) cycles.

Studying the relationship between cycle period and rotation period provides valuable insights into the characteristics of different types of stellar dynamos operating at different stellar ages. Brandenburg et al. (1998) and Saar & Brandenburg (1999) identified two distinct branches in the rotation period–cycle length ratio–activity (RCRA) diagram. Subsequent studies (Böhm-Vitense 2007; Boro Saikia et al. 2018) revealed two distinct branches in the rotation versus cycle-period diagram (henceforth referred to as the active and inactive branches), where the Sun is found to lie in the intermediate region between both branches. Böhm-Vitense (2007) proposed that the two sequences may correspond to distinct dynamo mechanisms, with one possibly occurring near the surface layer and the other operating deeper, near the interface between two zones. Metcalfe et al. (2016) concluded that our Sun which is lying between these two branches may be in a transitional evolutionary phase and that its magnetic cycle might represent a special case of stellar dynamo theory. Metcalfe & van Saders (2017) suggested that the cycle periods grow along the two sequences as stars spindown with the age, but at a critical Rossby number (Ro~2) the surface rotation rate change more slowly (van Saders et al. 2016) while the cycle gradually grows longer before disappearing. However, recent studies (Lehtinen et al. 2016; Boro Saikia et al. 2018; Olspert et al. 2018) have raised the doubts about the existence of the active branch. They found stars can lie in the intermediate region where our Sun resides, suggesting it may not be unique. Boro Saikia et al. (2018) suggested that the Sun's position in the cycle period–rotation diagram indicates that it is not an outlier and, thus, that the solar dynamo is likely representative of a common type of stellar dynamo. Therefore, comparing activity cycles of stars with different masses at various evolutionary stages is crucial for understanding how changes in their internal structures influence magnetic activity generation, particularly in the context of the solar dynamo.

This study is specifically focused on investigating the photometric activity cycles in fast-rotating stars, given the limited detection of such cycles in young stars till date. To derive activity cycles, we have combined photometric time-series data from different surveys. A key motivation is to use precise photometry, especially from *Kepler*, to detect activity cycles in a comprehensive sample of fast-rotating stars, especially in young Sun-like stars. This study is organised as follows. In Section 2 we describe our target sample and the methodology used to derive photometric activity cycles. In Section 3 we discuss the fitting of the combined time-series photometry from the different surveys to determine cycle periods. In Section 4, we present our results and analyze the relation of the derived activity cycle period with rotation. Followed by a discussion in Section 5 on how the activity cycles depend on different stellar parameters in the context of our Sun, which is also one of the goal of this study. Finally, we offer our conclusion in Section 6.

## 2 DATA AND OBSERVATIONS

### 2.1 Target Sample

This study focuses on analysing a subset of stars from our previously published catalogue of BY Draconis (BY Dra) variables (Chahal et al. 2022). BY Draconis are fast-rotating FGKM-type main-sequence stars which exhibit enhanced chromospheric emission (de Grijs & Kamath 2021). BY Dra stars also exhibit strong starspot modulation, for instance as observed in Zwicky Transient Facility (ZTF) light curves. The ZTF survey scans the entire northern sky every two nights (Bellm et al. 2018; Graham et al. 2019). The data release used in this paper, ZTF DR20 (Masci et al. 2018), covers the total observational span from March 2018 to October 2023. ZTF photometry is provided in the $g$ and $r$ bands, with a uniform exposure time of 30 s per observation.

Within the BY Dra subset, we identified a sample of stars exhibiting long-term photometric variations within the $\sim 5$ year ZTF DR20 time-series data. However, we found a significant fraction of stars exhibiting a single photometric cycle longer than the coverage of the ZTF DR20 observations. Hence, to detect longer activity cycles, we also included non-overlapping time-series observations from the *Kepler* Space Telescope and ASAS-SN. Consequently, we restricted our focus to sources in our ZTF-based BY Dra catalogue (78,954; Chahal et al. 2022) which were also observed by *Kepler* at a 30-minute cadence, or more frequently. We obtained *Kepler* Full Frame Images (FFIs) of 387 BY Dra and 160 RS Canum Venaticorum (RS CVn) sources (taken from Chen et al. 2020). RS CVn candidates are binaries with an F- to K-type giant or subgiant primary, which exhibit enhanced chromospheric activity (de Grijs & Kamath 2021). The *Kepler* FFIs are discussed in Section 2.3.1. However, for a significant fraction of BY Dra stars, standard *Kepler* light curves are not available (although we can access their *Kepler* FFIs). These will be examined in follow-up studies. The *Kepler* observations were obtained between 2009 and 2013.

Additionally, we also retrieved All Sky Automated Survey for SuperNovae (ASAS-SN) observations for ~150 stars with *Kepler*–ZTF data from the ASAS-SN Sky Patrol photometry database (Shappee et al. 2014; Hart et al. 2023) and from the ASAS-SN Catalog of Variable Stars (Jayasinghe et al. 2018, 2020, 2021), which contains observations in the $V$ band from ~2013 to 2018. The ASAS-SN observations also include $g$-band measurements, which we found to exhibit higher scatter compared to the ZTF $g$-band observations. Consequently, the variability features are not as prominent in ASAS-SN $g$-band measurements as in the ZTF $g$-band data. We opted to fit the combined Kepler–ASAS-SN ($V$-band)–ZTF ($g$- and $r$-band) data. Due to the significant wavelength overlap among these bandpasses, the combined fitting does not yield substantially different amplitudes. The combined photometric time series provides a baseline of ~14 years, which is hence suitable for detecting longer activity cycles, i.e., $P_{cyc} > 4$–5 years. We note that our target sample contains a subset of fast-rotating stars, i.e., mostly with $P_{rot} \leq 10$ days.

To include stars with longer periods, we incorporated the data set from Montet et al. (2017), which is based on a *Kepler* sample of rotational variables characterised by McQuillan et al. (2013). This data set comprises ~360 Sun-like stars observed with *Kepler* for which ZTF light curves are available. Consequently, out of an initial sample of 897 stars (comprising 387 BY Dra, 160 RS CVn and 360 Sun-like stars) with observations from ZTF and *Kepler*, ~25–30 per cent also have ASAS-SN observations. In this combined data set, we identified prominent activity cycles in 138 G–K-type main-sequence stars. The first set of 70 stars used data from *Kepler*–ASAS-SN–ZTF and the second set of 68 stars used data from *Kepler*–ZTF (see Section 3 for details). Additionally, we identified activity cycles in 25 RS CVn candidates. We note that several stars in our initial sample exhibit photometric variation in either one of the data sets, particularly in their *Kepler* or ZTF data. However, we focused on sources exhibiting a strong peak, corresponding to the cycle period in the power spectrum, and visually inspected those objects where the photometric fit effectively matched at least two of the data sets (*Kepler*, ASAS-SN or ZTF). For a detailed discussion as regards the selection of our sources and the detection of activity cycles, refer to Section 3. All identified sources are listed in Tables B1 (main-sequence stars) and B2 (RS CVn).

### 2.2 Photometric Observations

To illustrate the range covered by our target sample, we constructed a *Gaia* colour–magnitude diagram (CMD): see Fig. 1. Our sample predominantly consists of 138 G- and K-type stars and 25 RSCVn candidates. We calculated the absolute magnitudes of our sample stars using photogeometric distances derived from *Gaia* parallaxes, along with their colour and apparent magnitudes, employing the probabilistic approach by Bailer-Jones et al. (2021). The CMD in Fig. 1 shows our target sample of main-sequence (black data points) and RS CVn candidates (red data points), as well as a catalogue of BY Dra variables (grey data points).

From a target sample of 138 stars, 79 Sun-like stars were from the sample of Montet et al. (2017), and a very few of them lie above the main-sequence branch, possibly due to errors in distance estimation (see Fig. 1). Figure 3 of Chahal et al. (2022) illustrates a bifurcation in the CMD of our BY Dra catalogue (grey data points in Fig. 1). Chahal et al. (2022) argued that this bifurcation is most likely caused by the presence of binary companions, as indicated by the extended grey area above the main-sequence region in Fig. 1. Notably, most of our main-sequence target sample does not lie along the observed bifurcated branches. This suggests an absence of photometric binary systems in our sample (see Fig. 1).

Binarity check. To assess the effects of binarity in our target sample, we employed multiple diagnostic methods. First, we compared radial velocity (RV) measurements from LAMOST and the *Gaia* Radial Velocity Survey (RVS) (Gaia Collaboration et al. 2023a; Recio-Blanco et al. 2023; Das et al. 2025) for 138 stars, 58 of which had LAMOST RVs and 25 possessed multi-epoch observations. The strong correlation between LAMOST and *Gaia* RVs, despite being covered by different observation epochs, supports their classification as single stars. Multi-epoch LAMOST spectra (2012–2017) for 25 stars revealed RV variations within the typical 4—8 km s$^{-1}$ uncertainty range, except for two stars showing $\Delta$RV $\geq$ 10 km s$^{-1}$, suggesting potential binarity. Additionally, we cross-matched our stars with the *Gaia* DR3 Non-Single Star (NSS) catalogue (Gaia Collaboration et al. 2023b) and found no matches, indicating that there are no confirmed binaries based on *Gaia*'s criteria. We further examined the Renormalized Unit Weight Error (RUWE) values, which could reveal unresolved binaries through astrometric excess noise (Stassun & Torres 2021). Only 16 stars had RUWE $\geq$ 1.1, with two exceeding the binary threshold (RUWE $\geq$ 1.4); these have been flagged as potential binaries in Table 1. We also evaluated the single-star probabilities using *Gaia* DR3 parameters (DSC-Combmod, DSC-Specmod, and DSC-Allosmod) (Gaia Collaboration et al. 2016, 2023a), with all stars showing high probabilities (close to unity) of being single stars, except for one flagged source. We have flagged two sources with high RV variations, along with three identified based on RUWE and *Gaia* probabilities (five in total), as binary candidates in the Fit Quality column of Table 1 and have been excluded from the final analysis.

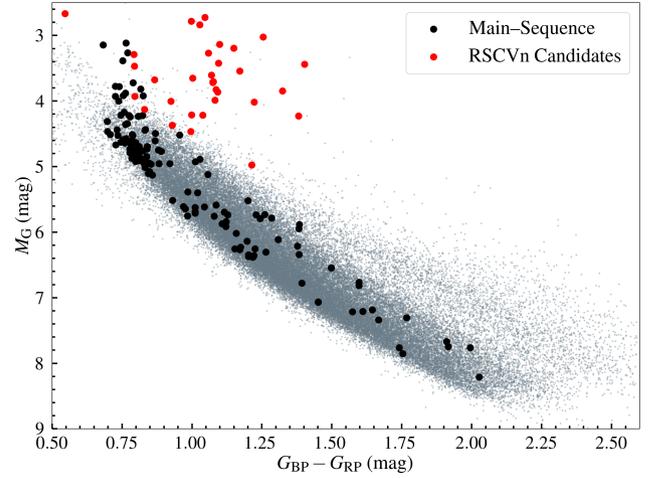

**Figure 1.** Absolute *Gaia* G-band magnitude versus $(G_{BP} - G_{RP})$ colour diagram. Grey data points represent entries from the BY Dra catalogue compiled by Chahal et al. (2022). Black and red data points denote, respectively, main-sequence stars and RS CVn candidates in our sample.

The photometric observations collected from *Kepler*, ASAS-SN and ZTF used different wavelength filters, resulting in different variability amplitudes in each waveband. In addition, the combined light curves from these data sets encompass starspot modulation, multiple cycles, etc., which can be challenging to isolate. Moreover, even long-term observations of the solar cycle have revealed variations in the amplitude across each cycle (Krivova et al. 2006; Solanki et al. 2006, e.g.). In fact, Shapiro et al. (2016) demonstrated that the Sun's amplitude of variability also varies as a function of wavelength. Hence, to reduce the impact of shorter-time-scale fluctuations and highlight long-term trends, we opted to bin the data over extended time intervals, defined by $n \times P_{\rm rot}$. We will discuss the method in detail in Section 2.3.

### 2.3 Methodology

The impact of rotational modulation caused by starspots or faculae can significantly affect long-term trends. Mathur et al. (2014) introduced the $S_{\rm ph}$ index, derived by binning time-series data with multiples of $5 \times P_{\rm rot}$ and computing the standard deviation within each bin. The resulting average standard deviation, denoted by $\langle S_{\rm ph} \rangle$, serves as a measure of a star's photospheric activity. Notably, fig. 1 in Mathur et al. (2014) demonstrated that the standard deviation of solar data, obtained through binning, exhibits cyclic variations similar to the solar cycle. Similarly, Reinhold et al. (2017) detected photometric variability by binning *Kepler* data over a single day and computed the $R_{\rm var}$ index, i.e., the difference between the fifth and 95th percentiles of the relative flux. The activity cycles were determined by fitting a sinusoidal function to the $R_{\rm var}$ values.

Drawing from prior studies on binning photometric data to discern activity cycle periods, we employed a similar approach. We binned our time-series data by $n \times P_{\rm rot}$, where $P_{\rm rot}$ denotes the rotation periods derived from the *Kepler* light curves, and computed the standard deviation and mean flux within each bin. However, instead of adhering strictly to $n = 5$, as suggested by Mathur et al. (2014), we adjusted $n$ based on the stellar rotation period. This is because a fixed $n = 5$ might significantly limit the number of data points, particularly for stars with rotation periods exceeding 15 days. This would compromise the accuracy of our activity cycle period determination. Conversely, for stars with rotation periods shorter than 2 days, binning by a factor of five could allow shorter-term variability signatures to remain hidden within long-term trends. We note that the cycle periods in our study were derived by fitting the mean flux values within the respective bins, in contrast to Mathur et al. (2014), who used the standard deviation within the bins to determine the cycle periods. In our analysis, the standard deviations are only used to examine possible spot or faculae dominance.

To address these considerations, we adapted a new binning strategy tailored to this study: for stars with longer rotation periods ($P_{\rm rot} \geq$ 10 days), we implemented a binning size of $P_{\rm rot}$, i.e., ($n$ = 1), to preserve data resolution, whereas for stars with shorter rotation periods ($P_{\rm rot} <$ 10 days), we adjusted the bin size within a 10–20 day bin width. The bin range was decided upon after experimentation with various bin sizes to identify the optimal cycle period (for example, stars with $P_{\rm rot}$ = 2, 5, 10, 15, etc., were binned with $n$ = 8, 2, 1, 1, etc., respectively). Subsequently, we derived the mean fluxes and standard deviations within these bins, which we show as a function of time in Figs 2, 3 and 4. The variability in mean flux reflects the transition of stars between their maximum and minimum states, while the standard deviation indicates the amplitude of the rotational modulation during these phases.

To find the optimum periods, we performed Fourier analysis of the time-series data. The Lomb–Scargle periodogram (Lomb 1976; Scargle 1982) is the most widely used tool for measuring quasi-periodic rotation and activity cycle periods. Boro Saikia et al. (2018) employed Generalised Lomb–Scargle Periodogram (GLS) tools (Zechmeister & Kürster 2009) to identify activity cycle periods. We also applied the GLS tool to detect the strongest peak, which is assumed to be indicative of the activity cycle period. We used the *Python–PyAstronomy* GLS package to estimate and fit the cycle periods using the best-fitting sine curve (see Fig. 2). In addition, this tool facilitated the derivation of a False Alarm Probability (FAP), representing the likelihood of mistakenly detecting a peak. It is used to estimate the likelihood that, in a power spectrum resulting from random noise, at least one of the $M$ independent power values will reach or exceed a defined threshold: see Table B1 (for details, see the *Python–PyAstronomy* GLS package). We compared the activity cycle periods derived using our method for the ZTF data with those obtained from *Kepler* FFIs by Montet et al. (2017): see Fig. A1 in Appendix A. We observed a clear one-to-one correlation between the cycle periods determined from the ZTF and *Kepler* data sets (see Fig. A1), except for two sources. Consequently, combining both data sets can yield more robust estimates of the cycle periods, particularly for longer cycles.

### 2.3.1 *Kepler* Full Frame Images

Long-term signals in *Kepler* light curves, lasting approximately 50 days or more, attributed to brightness variations linked to magnetic cycles, present challenges in data analysis. These signals are deliberately filtered out by the data processing pipeline of *Kepler* (Gilliland et al. 2011). However, the *Kepler* FFIs offer a solution to recover these long-term brightness variations (Montet & Simon 2016). The *Kepler* spacecraft captured roughly one FFI per month throughout the primary mission from 2009 to 2013. Employing the *f3* software package (Montet et al. 2017), we conducted photometry on the *Kepler* FFIs, thus enabling detection of changes in stellar brightness over extended time spans.

The flux derived from FFIs contains contributions from starspot/faculae modulation. Given the sparse nature of FFI data points, binning proved ineffective. Therefore, we fitted the long-cadence *Kepler* light curves with multiple fourth-order Fourier functions of rotation periods, focusing on the narrow regions around each *Kepler* FFI data point. This approach ensured that the fits were not influenced by starspot or faculae evolution. We then subtracted the estimated flux from the *Kepler* light curve at each FFI data point from the FFI fluxes, effectively removing the contribution of rotational modulation. This process isolates long-term trends in the *Kepler* FFIs. Additionally, we computed the standard deviations for the *Kepler* light curves within $n \times P_{\rm rot}$ (which provides an estimate of the rotational modulation) for each *Kepler* FFI data point: see Figs 2 and 3.

## 3 PHOTOMETRIC STELLAR ACTIVITY CYCLES

It is well-established that cooler stars like our Sun display chromospheric (Boro Saikia et al. 2018; Olspert et al. 2018) and photometric activity cycles (Reinhold et al. 2017). Our target sample (see Section 2.1) primarily contains fast-rotating stars G–K-type stars, with Sun-like stars taken from Montet et al. (2017). We derived the activity cycles for these stars using the methodology described in Section 2.3. We found 138 main-sequence stars with strong peaks in their power spectra corresponding to the cycle periods, which were visually inspected and are listed in Table B1. Stellar cycles are complex and light curves often deviate from a simple sinusoidal shape, as seen in the Sun's asymmetric rise and fall or its double-peaked maxima from hemispheric differences (Hathaway 2010). However, driven by our binning strategy and multi-wavelength fitting approach, we have modelled the data using a simple sinusoidal function.

The combined time series from the *Kepler*-FFIs–ASAS-SN–ZTF surveys spans approximately 14 years. Following our binning procedure, we calculated the mean flux and standard deviation within each of these bins as a function of time. We marked the strongest peak in the GLS power spectrum as the activity cycle period. To assess our period accuracy across different bands, we fitted the cycles using both the ZTF $g$ and $r$ bands (see Fig. 2, where the left- and right-hand panels correspond to the ZTF $g$ and $r$ bands, respectively). We retained sources where both bands exhibited similar cycle peaks in their power spectrum (see subsection 3.1). In this section, we discuss the fitting behaviour of the cycle periods within our sample for the different data sets.

### 3.1 *Kepler*–ASAS-SN–ZTF Cycles

The first set contains 70 sources with *Kepler*–ASAS-SN–ZTF data: 62 from ASAS-SN Sky Patrol photometry database (Shappee et al. 2014; Hart et al. 2023) and 8 from ASAS-SN catalogue of variable stars (Jayasinghe et al. 2021). These sources displayed cyclic variations corresponding to atleast one strong peak in the power spectrum and fitting effectively at least two of the data sets (*Kepler*, ASAS-SN or ZTF). For example, the fits in Figs. 2 & 3 fit the *Kepler* and ZTF data sets effectively, but the ASAS-SN data exhibit significant scatter. We observed slightly different variability amplitudes across different data sets (*Kepler*–ASAS-SN–ZTF), which is expected because of the different passbands employed (see Fig. 3). The ASAS-SN data showed the largest scatter in comparison with *Kepler* and ZTF (see Figs 2, 3 and 4). Based on visual examination, we found that the ZTF $g$ band seems most sensitive to cyclic photometric variations compared with the other filters in this data set.

We show example sources with their power spectra in Figs 2 and 3, and additional sample fits in Figs A2 and A3 in Appendix A.

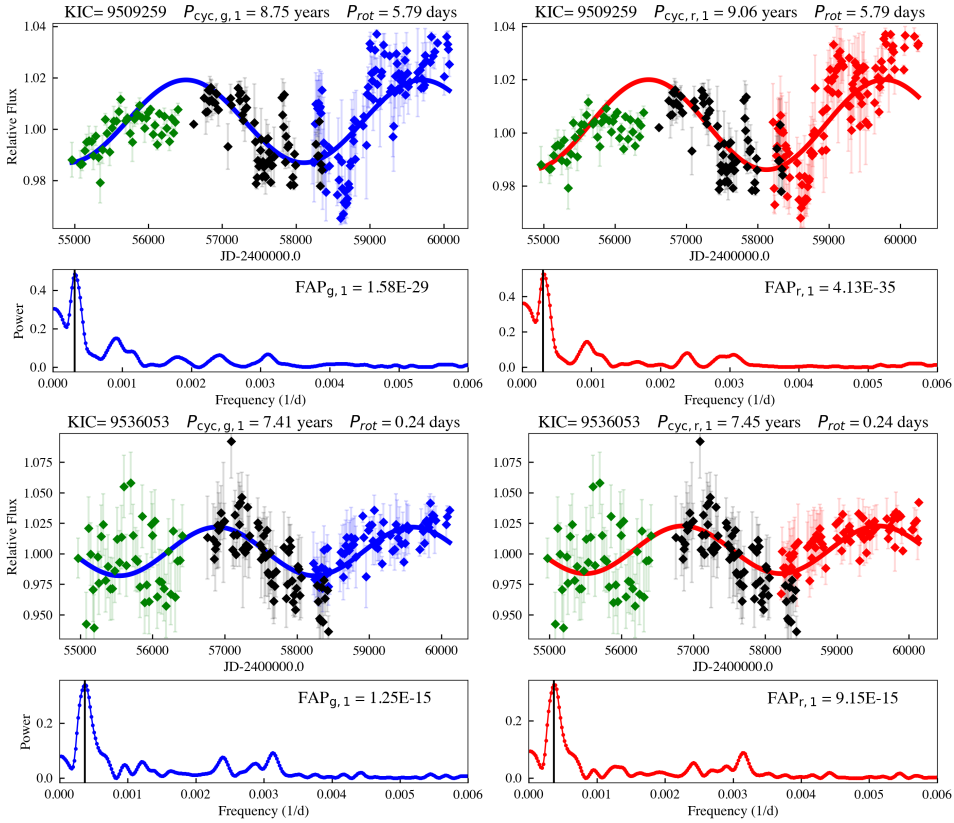

**Figure 2.** Example of photometric cycle fitting to the combined *Kepler* (green)–ASAS-SN (black)–ZTF data for two sources – (top) a Sun-like star and (bottom) an RS CVn candidate – featuring the ZTF *g* band (blue) in the left-hand panels and the *r* band (red) in the right-hand panels. The error bars on the photometric fits represent the standard deviation within each bin. Below each photometric data panel, we have included the power spectrum displaying the peak corresponding to the best-fitting period. The *Kepler* Input Catalogue (KIC) ID, the best-fitting cycle periods and the rotation periods of the sources are annotated within their respective panels. They are also listed in Table B1.

Figure 2 presents two such sources, where the left- and right-hand panels fit the data covering the *Kepler*–ASAS-SN–ZTF *g* and *r* bands independently. We have excluded sources that do not exhibit cycle period peaks within ±1 year in the power spectrum of *Kepler*–ASAS-SN–ZTF *g* and *Kepler*–ASAS-SN–ZTF *r* bands. This exclusion was primarily applied to stars lacking intermediate ASAS-SN observations, to ensure consistency in cycle-period measurements between the two ZTF bands. Note that all targets in our sample have fractional errors below 10 per cent, with an average fractional error of ~3–5 per cent: see Table 1. This difference between periods might be owing to the dependence of the variability amplitude on wavelength, also observed for the Sun (Shapiro et al. 2016). We found that approximately 30–40 per cent of our initial sample exhibited such inconsistencies in their cycle peaks between the ZTF *g* and *r* bands. We present the list of derived and best-fitting parameters, including the FAP, variability amplitude and phase correlation in Tables B1 and B2. The fits for all sources are available online[1].

### 3.2 *Kepler*–ZTF data

Additionally, we also detected prominent activity cycles in our second set comprising of 68 sources with *Kepler*-FFI–ZTF data for which ASAS-SN data are not available using the methodology described in subsection 2.3. Although in some cases photometric variability may not be prominent in the *Kepler* FFIs, it is noticeable in the ZTF *g* and *r* bands. We found good agreement among cycle periods for sources with available ASAS-SN data when fitting them both with and without the ASAS-SN data. This confirms that the cycle period accuracy is still good, even if we do not have intermediate-period ASAS-SN data.

[1] https://doi.org/10.5281/zenodo.15336130

In the right-hand panel of Fig. 3, we show a selection of stars with *Kepler*–ZTF data exhibiting single prominent peaks in their power spectra, with the average $FAP_g$ of $10^{-5}$ of all the *Kepler*–ZTF fits. To ensure accurate determination of the cycle period, we have fitted the data for both the ZTF *g* and *r* bands independently (similar to those shown in the left- and right-hand panels of Fig. 2, respectively). We excluded sources where the strong peak was not consistent between both bands. In Fig. 3 we only show fits for the ZTF *g* band, but fits for all sources with both *g*- and *r*-band data are available online[1]. In several instances, more than one dominant peak was observed in the power spectrum. This is discussed in Section 3.2.1.

#### 3.2.1 Activity – cycle classifications

In ~22 per cent (30/138) of our sample we observed significant multiple peaks in the power spectrum, notably two peaks in both the ZTF *g* and *r* bands. Often, one peak appears to be a harmonic of the other, thus making it challenging to ascertain the true cycle peak. Consequently, through visual inspection, we obtained $P_{cyc,2}$ for stars with double cycle period peaks, listed in Tables B1 and B2. A sample of such fits with double peaks in their power spectra is shown in Fig. A3 in Appendix A; all fits are available online[1]. We observed that the second or third peak is often approximately twice the period of the first peak, or half of it in some cases.

Recent work by do Nascimento et al. (2023) has shown that 18 Sco, a younger Sun-like star, exhibits a shorter 6.9-year cycle equivalent to the 11-year solar Schwabe cycle, as well as a longer, ~15-year cycle analogous to the 22-year Hale magnetic cycle of the Sun. Note that double peaks can also arise in stars that cross the spot-to-faculae dominance boundary. As a star transitions between spot- and faculae-dominated phases, changes in brightness can mimic a reversal in activity, potentially producing two distinct cycles while the overall activity continues to increase. This effect might be more pronounced for K-type stars where the spot-to-faculae transition is not known, whereas in Sun-like stars, this transition typically occurs around a rotation period of $P_{rot} \sim 20$–25 days. Alternatively, Olspert et al. (2018) have proposed that these double cycles can also result from a single, quasi-periodic cycle. It is possible that these double-peak harmonics might be caused by missing data in the interval between the *Kepler* and ZTF observations. However, we also observed double peaks in combined *Kepler*–ASAS-SN–ZTF fits.

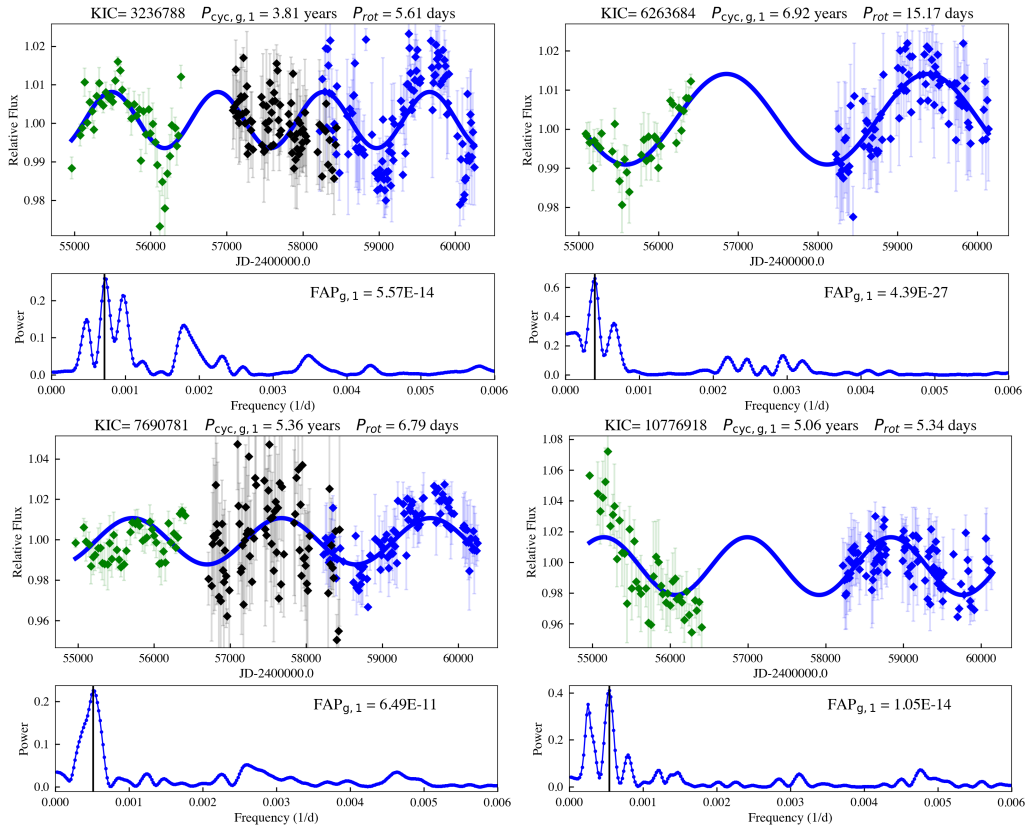

**Figure 3.** Examples of photometric cycle fitting to the combined *Kepler* (green) – ASAS-SN (black) – ZTF *g*-band (blue) (first-set) data in the left-hand panels and *Kepler* (green) – ZTF *g*-band (blue) (second-set) data in the right-hand panels. The error bars on the photometric fits represent the standard deviation within each bin. Below each photometric data panel, we have included the power spectrum displaying the peak corresponding to the best-fitting period, along with the FAP of the cycle peak. A larger sample of fits is included in Fig. A2 in Appendix A.

In Appendix B we will further discuss the possibility of stars having two apparent cycles or alias periods. Meanwhile, we plotted $P_{cyc,1}$ and $P_{cyc,2}$ in Fig. B1. Additionally, we examined the phase-folded light curves for stars with multiple periods and found no significant differences in the shapes of their folded light curves. Although there are two peaks in the power spectrum, the average FAP of the strongest peak period is $\sim 10^{-4}$ for most of our sources. Therefore, we considered the period with the lowest FAP as representative of the most probable activity cycle period. A sample of stars with multiple peaks in their power spectra is shown in Fig. A3, where we fitted the light curves with the period corresponding to the lowest FAP peak. We list the cycle periods, the corresponding FAPs and the best-fitting parameters of the main-sequence stars in Table B1 and the equivalent data for the RS CVn candidates in Table B2.

In addition to identifying single and multiple cycles, we have encountered instances where the photometric variation is more pronounced in one of the wavebands. Accordingly, we have categorised the sources based on the relative peaks in the power spectrum, the FAP of the strongest peak and a visual fit inspection into 'good' and 'average' activity cycle fits denoted as 'Fit Quality' in Tables B1 and B2. We found that $\sim 73$ per cent (101/138) of sources with good cycle fits have $\langle FAP_g \rangle \sim 10^{-6}$, whereas the remaining $\sim 27$ per cent (27/138) displaying average cycle fits have $\langle FAP_g \rangle \sim 10^{-3}$. The activity cycles of stars classified as 'good' have at least one or two cycle crests and troughs fitted within the photometric time-series data. The overall average FAP of our sample is approximately $10^{-4}$, which is better than in previous photometric studies (Suárez Mascareño et al. 2016; Irving et al. 2023), due to precise photometry from *Kepler* and ZTF.

### 3.3 Estimating Starspot–Faculae dominance

The variations in magnetic activity define the stellar activity cycles. Hence, the variation in photospheric activity has been directly linked to long-term brightness fluctuations (Mathur et al. 2014; Salabert et al. 2016). The standard deviation of the observational data serves as a proxy for the photospheric activity, with indices like $R_{per}$ and $S_{ph}$ commonly used for this purpose (Mathur et al. 2014; McQuillan et al. 2014). Thus, analyzing the standard deviation within each data-set bin to explore its correlation or anti-correlation with any photometric variations. This relationship can provide insights into whether a star's surface is dominated by the presence of starspots or faculae. For instance, Montet et al. (2017) investigated the correlation between *Kepler* FFI photometric variability and the *Kepler* light curve's standard deviation ($S_{ph}$). Similarly, Reinhold et al. (2019) examined the phase difference between chromospheric and photometric variability to determine whether a star's surface is dominated by faculae or starspots.

However, a higher standard deviation does not always correspond to the maximum activity phase. An increase in the number of spots or faculae across the stellar disc can lower the standard deviation (RMS) owing to a more uniform distribution. Similarly, shifts from axisymmetric to non-axisymmetric spot distributions can change the RMS without necessarily reflecting changes in the star's activity cycle phase. While multi-epoch spectroscopic observations of the Ca II H and K lines offer a more reliable estimate of spot or faculae dominance (Lockwood et al. 2007; Radick et al. 2018), such data are unavailable for our sample. Instead, we find clear evidence of spot and/or faculae evolution in some of the *Kepler* RMS data (see Figure 4), enabled by its high-precision photometry. This corresponds to maximum stellar activity phases when spot redistribution or evolution occurs. For these stars, RMS variations can effectively reflect whether a star is spot- or faculae-dominated. Recent studies have also used *Kepler* data to trace spot distribution and evolution (Namekata et al. 2020; Xu et al. 2021). We also acknowledge the limitations of RMS in distinguishing spot- and faculae-dominance, given the lack of multi-epoch spectral observations of Ca II H&K, H$\alpha$, Ca II IRT etc, tracking spot evolution in *Kepler* RMS remains the most viable approach.

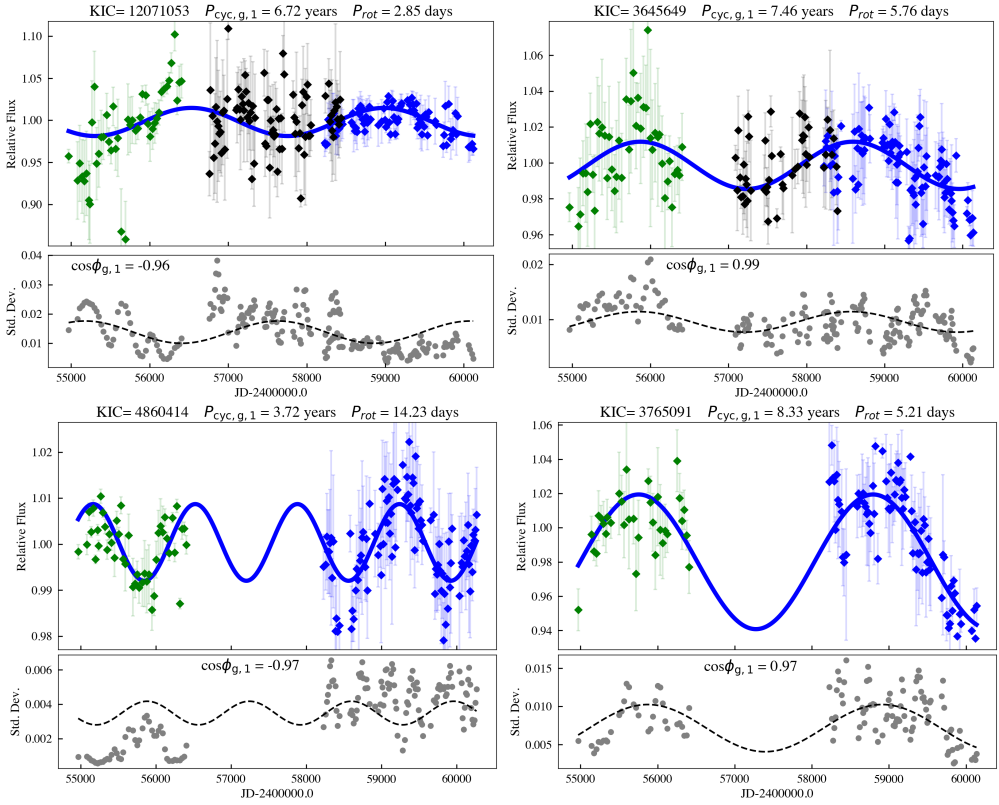

**Figure 4.** Example of fitting the cycle period to the photospheric activity (standard deviation). The photometric cycle fits to the combined *Kepler* (green)–ASAS-SN (black)–ZTF *g*-band (blue) (first-set) data are depicted in the top panels and to the *Kepler* (green)–ZTF *g*-band (blue) (second-set) data in the bottom panels. The bottom panels of each source show the standard deviation (grey) (RMS) of each photometric bin fitted versus the derived activity cycle periods. The phase difference between the sine-wave representation of the photometric data and their photospheric activity offers insights into the (anti-)correlations. The cosine of the phase difference provides a quantitative measure of whether the star is faculae- ($\cos\phi = +1$) or starspot-dominated ($\cos\phi = -1$).

We conducted an analysis of the phase differences between the mean flux variations and standard deviations. We fitted the standard deviation with a sine curve using the same cycle period which was derived from the mean flux variations, and determined the phase difference ($\phi$) between the two sine waves. Figure 4 illustrates examples of stars fitted with sine wave-like cycle periods. Solid lines indicate fits to the mean flux, whereas dashed lines are used to fit the standard deviation. The cosine of the phase difference between the two fits indicates a correlation or anti-correlation between the two sine waves. A strong correlation or anti-correlation, combined with the evidence of spot or faculae evolution in the *Kepler* RMS data, suggests that maximum flux either correlates or anti-correlates with the maximum stellar activity phase. For instance, the left-hand panels of Figure 4 shows pronounced spot/faculae evolution in the *Kepler* RMS data, with the cycle period strongly anti-correlated with the spot evolution phase. This indicates that such evolution primarily occurs during periods of maximum activity, suggesting that the star is spot-dominated. The significant decrease in the *Kepler* RMS afterwards reflects the emergence, disappearance and redistribution of spots on the stellar surface. As the star transitions into a phase of minimum activity, both the RMS and the spot evolution trends reduce. By contrast, the right-hand panels of Figure 4 displays spot/faculae evolution that correlates with brightness variations. This correlation between the phase of maximum activity and flux changes suggests that the star is faculae-dominated.

Therefore, we visually inspected each source and classified only those displaying clear trends of spot or faculae evolution in the *Kepler* RMS data, indicating strong correlations or anti-correlations with flux variations (see Fig. 4). In some cases, the ASAS-SN data showed relatively higher standard deviations compared with ZTF or *Kepler*, which dominated the fits. We list the cosine of the phase difference between the two waves in both the *g* and *r* bands in Tables B1 and B2 for all sources. However, only 27 sources met our classification criteria (18 spot-dominated and 9 faculae-dominated), as listed in Table 1. This classification applies to sources with $|\Delta\cos\phi| < 0.5$ between the *g* and *r* fits and $|\cos\phi_{g,r}| > 0.5$, provided their amplitudes are significant relative to the scatter in the data and they exhibit spot/faculae evolution in *Kepler* RMS, as determined through visual inspection. Sources that do not meet these criteria remain unclassified in Tables B1 and B2. All photospheric activity fits are available online[1].

### 3.4 Chromospheric Activity Indices

In addition to analysing activity cycles, we conducted a cross-match of our target sample with the low-resolution spectra from the Large Sky Area Multi-Object Fibre Spectroscopic Telescope (LAMOST; the Guo Shoujing Telescope) Data Release 9 (DR9; Cui et al. 2012). Among these spectra, we identified 50 sources with signal-to-noise ratios (SNR) of *g* filter greater than 30. From these spectra, we derived chromospheric activity indices based on H$\alpha$ line emission. Note that H$\alpha$ emission can be challenging to interpret owing to absorption from prominences (Meunier et al. 2022; Gomes da Silva et al. 2022), although this effect is primarily significant for M-type stars (Cram & Giampapa 1987). Additionally, we estimated the Ca II H and K indices using the ACTIN2 package (Gomes da Silva et al. 2018, 2021), similarly to Chahal et al. (2023). We found a strong one-to-one correlation between the H$\alpha$ and Ca II H and K indices. Given this correlation, the absence of M-type stars in our target sample (which contains only two M-type stars, without LAMOST spectra), and the almost double SNR of the H$\alpha$ measurements, we have adopted the H$\alpha$ indices as our primary diagnostic. For a comprehensive discussion of the derivation of $R'_{H\alpha}$ and $R'_{HK}$ indices, we direct readers to our earlier study (Chahal et al. 2023), which focused on deriving chromospheric activity indices from the LAMOST spectra.

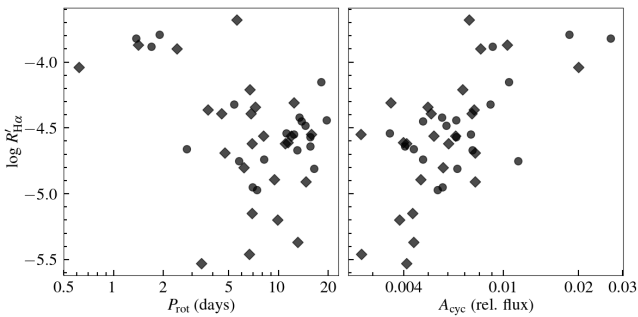

**Figure 5.** Chromospheric activity indices, $\log R'_{H\alpha}$, as a function of (left) rotation period ($P_{rot}$) and (right) activity cycle amplitude ($A_{cyc,g,1}$) for the sources listed in Table B1. The diamonds and circles represents data belongs to the first-set and second-set, respectively.

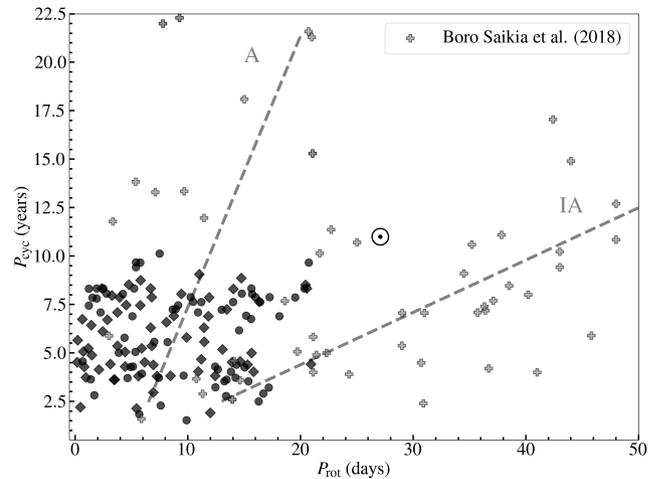

**Figure 6.** Activity-cycle period ($P_{cyc,g,1}$) as a function of rotation period ($P_{rot}$) for the stars in Table B1. The black diamonds and circles represent cycle periods derived from fitting, respectively, first-set (*Kepler*–ASAS-SN–ZTF) and second-set (*Kepler*–ZTF), which are included in Table B1. The grey plus symbols were taken from Boro Saikia et al. (2018), which were derived using the chromospheric activity indices (*S*-indices). The grey dashed lines show the active and inactive branches according to Böhm-Vitense (2007). The Sun is shown as ⊙.

We have plotted the derived chromospheric activity indices in relation to the rotation period and cycle amplitude in Fig. 5. The chromospheric activity indices exhibit trends consistent with our previous study, which encompassed a larger sample (Chahal et al. 2023). Additionally, we observe a notable positive correlation between the chromospheric activity indices and the activity cycle amplitudes (see Fig. 5, right). However, the observed scatter might be owing to single-epoch observations of the chromospheric activity indices, since chromospheric activity is expected to vary with activity cycle. The scatter observed for some K-type stars may also result from their transition between spot- and faculae-dominated phases, as this transition can reduce the cycle amplitude. The positive correlation implies that the higher chromospheric activity is associated with stronger magnetic activity cycles. A more detailed discussion of the derived activity indices in relation to the corresponding cycle periods is provided in Section 4.

## 4 RESULTS

Previous studies (Böhm-Vitense 2007; Brandenburg et al. 2017; Boro Saikia et al. 2018; Olspert et al. 2018) have demonstrated an observed relationship between activity cycle and rotation period, purportedly identifying two branches within stellar activity cycles. Here, we investigate the relationship between activity cycles and rotation periods based on the cycle periods determined in Section 3; see Table B1. We note that we have mainly detected cycle lengths that are less than two-thirds of the total data-set time span, i.e., $P_{cyc} \leq 10$ years. This upper limit ensures that at least one peak and trough could be adequately fitted with the available data, especially considering that half of the sources do not have observations in the interval between the *Kepler* and ZTF data sets. We classify our samples as exhibiting 'good' activity cycles when the data covers more than one cycle length, defined by either two peaks or two troughs. Furthermore, these stars demonstrate a distinct peak in their power spectrum with

$\langle FAP_g \rangle \sim 10^{-6}$. The average errors in their cycle periods are 4-5 per cent, listed in Tables B1 and B2.

Additionally, to ensure the cycle periods are not biased by the combination of different time series, we also independently estimated cycle periods using only Kepler FFIs and only ZTF data. Given the baselines of the Kepler FFIs (~4 years) and ZTF (~5 years), we restricted the comparison to stars with cycle periods under 5 years. The cycle periods derived from the *Kepler* FFIs and ZTF data show strong agreement with those obtained from combined photometry using *Kepler*–ASAS-SN–ZTF and *Kepler*–ZTF.

We note that only a subset of activity cycle fits are shown in Figs 2, 3 and 4, and in Appendix A, although all cycle fits can be accessed online[1]. Similarly, a sample of our tables is presented in Tables B1 and B2; full tables are also available online[1].

### 4.1 Activity Cycles in Main-Sequence Stars

To explore the relationship between rotation and activity cycle periods, we present our main-sequence stars in Table B1 and plot their activity cycle periods as a function of rotation period, similar to fig. 1 of Böhm-Vitense (2007) and fig. 9 of Boro Saikia et al. (2018). A decade-long debate has centered around the possible classification of stars into two branches (active and inactive) in the activity cycle period versus rotation diagram. To ascertain the reality of the active and inactive branches in our sample, we plot their activity cycle periods and rotation periods in Fig. 6. Figure 6 shows distinct cycles marked in black. In addition, we overlay cycle periods from Boro Saikia et al. (2018), given that our sample predominantly falls within the fast-rotating regime (black data points). A similar plot, combined with multiple cycles, is provided in Appendix B; see Fig. B2.

In Fig. 6, we observe that the black data points diverge from the distinct active and inactive branch trends (grey dashed lines in Fig. 6) identified in previous studies (Böhm-Vitense 2007; Boro Saikia et al. 2018). Instead, our data set exhibits considerable scatter, thus suggesting that the existence of two distinct branches, particularly the active branch, may not hold true for young stars. However, including slow-rotating stars can provide better constraints on the existence of both branches. A similar kind of spread in the active branch was also observed by Boro Saikia et al. (2018) (see the grey data points in Fig. 6). However, we note that while the activity cycles in Boro Saikia et al. (2018) were derived using chromospheric indices from spectra, we derived activity cycles using combined photometry. Our results indicate that a significant fraction of stellar activity cycles could lie well above the active branch. Moreover, we did not discern any linear trends among stars situated closer to the active branch. It is plausible (based on Fig. 6) that several stars are located in the intermediate region between the proposed active and inactive branches, as further discussed in subsection 5.2.2. Similar scatter around the active branch is also seen for stars with multiple cycle periods: see Fig. B2 in Appendix B.

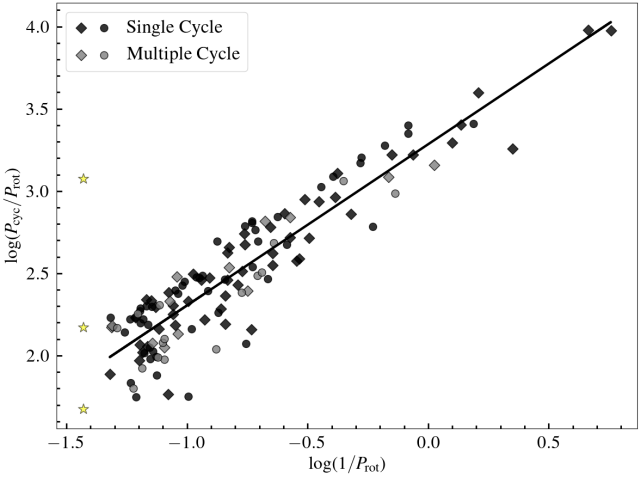

**Figure 7.** Relationship between the observed rotational ($P_{rot}$) and cycle periods ($P_{cyc,g,1}$). Black data points represent stars with a single cycle ($P_{cyc,g,1}$), while grey points indicate stars with multiple cycles (with $P_{cyc,g,1}$ & $P_{cyc,g,2}$). The diamonds and circles represents data belongs to the first-set and second-set, respectively. Solar data are shown as yellow stars (Gliessberg, Schwabe and 3–4-year cycles). The fit was done for both stars with single and multiple cycles. See the text for details.

To further investigate the relationship between the length of the magnetic cycle and the rotation period, we have plotted $P_{cyc}/P_{rot}$ versus $1/P_{rot}$ in Fig. 7. This parameter space has been investigated extensively by Baliunas et al. (1996), Oláh et al. (2009, 2016) and Suárez Mascareño et al. (2016). It is anticipated that the cycle period scales as $\sim D^l$, where $l$ represents the slope of the relationship and $D$ denotes the dynamo number, which is a measure of the efficiency of the dynamo mechanism. Deviations from $l \approx 1$ suggest a correlation between the cycle length and rotation period. For instance, slopes of 0.74 and 0.81 were reported by Baliunas et al. (1996) and Oláh et al. (2016), respectively. However, Suárez Mascareño et al. (2016) observed a slope of $0.99 \pm 0.07$, indicating no correlation between these quantities. Nevertheless, they found a slope of $0.89 \pm 0.05$ specifically for main-sequence FGK stars, suggesting a weak correlation in their data.

We have fitted $P_{cyc}/P_{rot}$ versus $1/P_{rot}$ in Fig. 7, similarly to Baliunas et al. (1996). In our sample of fast-rotating stars, we found a slope of $1.01 \pm 0.01$. This indicates no discernible correlation between both quantities, which is consistent with the conclusion of Suárez Mascareño et al. (2016). For stars with multiple cycles, we have fitted the quantities in Fig. 7 with $P_{cyc,1}$. However, even with

$P_{cyc,2}$, we observe a slope of $1.04 \pm 0.01$ between the two quantities. If we remove the Sun-like stars taken from Montet et al. (2017), the slope changes to $0.95 \pm 0.01$, suggesting no, or perhaps a very weak, correlation. A similarly weak correlation has also been noted by Suárez Mascareño et al. (2016) and Oláh et al. (2016).

### 4.2 Starspot versus Faculae dominance

The photospheric activity cycles are either dominated by starspots or faculae, which is discussed in Section 3.3. Previous studies (Montet et al. 2017; Reinhold et al. 2019) have examined (anti-)correlations between brightness fluctuations and magnetic activity, providing insights into whether a star is starspot- or faculae-dominated, respectively (see subsection 3.3 for more details). We have similarly analysed the phase differences to determine whether the brightness fluctuations correlate or anti-correlate with photospheric activity. We have discussed the challenges of using RMS to identify spot-facula dominance in Section 3.3. We have considered sources only when the amplitude is visually significant and consistent among the *Kepler* and ZTF *g* and *r* bands. However, we note that the standard deviation in ASAS-SN is significantly larger in comparison with the ZTF and *Kepler* data. Therefore, we have carefully examined the fitting quality to classify the sample into faculae- or starspot-dominated; see Table B1.

In contrast to Montet et al. (2017) and Reinhold et al. (2019), we find a sample of fast-rotating K-type stars which are dominated by faculae (see black data points in Fig. 8). Montet et al. (2017) proposed a transition from starspot to faculae dominance around rotation periods of 15–25 days for Sun-like stars. Since we did not observe activity cycles for longer-period stars, our study cannot provide constraints on this starspot-to-faculae transition. However, we did identify activity cycles in 79 Sun-like stars from the sample of Montet et al. (2017). We observe seven starspot-dominated (grey) Sun-like stars ($T_{eff} \in 5500 - 5900$K) at shorter periods, consistent with Montet et al. (2017), as shown in the second panel of Fig. 8. Conversely, we observed 9 faculae-dominated (black) fast-rotating cooler stars. This suggests that while the transition from starspots to faculae may occur in Sun-like stars, as proposed by Montet et al. (2017) and Reinhold et al. (2019), that might not hold true for cooler stars. Additionally, we also find 4 K-type stars which are faculae-dominated, and these stars also found have dual cycle. One possibility can be that these might have slipped to the spot/faculae dominance boundary. However, if that were the case, their correlation coefficient ($\cos(\phi)$) should be below 0.5, whereas we found it between 0.8 and 1.

Shapiro et al. (2016) argued that the relative contribution of faculae increases for Sun-like stars at lower inclinations. In fast rotators, magnetic buoyancy may drive spots poleward, leaving equatorial regions dominated by faculae. We cross-matched our sample with Frasca et al. (2016) for inclination data but found that only a few sources, including faculae-dominated stars, have high inclinations. We plan to further investigate the dependence of starspot and faculae contributions to inclination in the future using medium- or high-spectral-resolution data. Additionally, we found some evidence suggesting that fast-rotating K-type stars dominated by faculae have lower metallicities compared with starspot-dominated stars, based on metallicity derived through LAMOST pipeline using low-resolution spectra. We aim to further investigate this dependence using medium or high-resolution spectra to better understand the relationship between metallicity and starspot or faculae dominance. We also note that while we have critically examined faculae evolution using the RMS of these stars, Ca II H&K variations may serve as a stronger indicator of faculae dominance, which we aim to explore in future work. Gully-Santiago et al. (2017) modelled spot coverage by fitting spectra with two-temperature model, Feinstein et al. (2021) also reported high spot coverage in chromospheric lines; we aim to refine this by estimating spot coverage in our sample with high-resolution spectra.

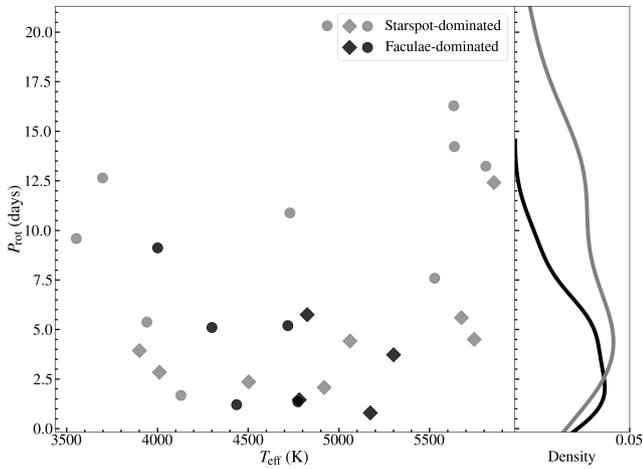

**Figure 8.** Rotation period versus effective temperature for stars classified as starspot- or faculae-dominated in Table B1. The diamonds and circles represent cycle periods derived from fitting the *Kepler*–ASAS-SN–ZTF (first-set) and *Kepler*–ZTF (second-set) data sets, respectively, which are listed in Table B1. The black and grey colours represent stars dominated by faculae and starspots, respectively, with their density distributions depicted in black and grey, respectively.

### 4.3 Activity Cycles in RS CVn Variables

In addition to investigating activity cycles in main-sequence stars, our study extends to RS CVn candidates, classified by Chen et al. (2020). These candidates represent chromospherically active binaries, typically comprising a primary evolved star with a G–K-type main-sequence companion. For our analysis, we focused on a subset of these candidates with initial sample of 160 sources with available *Kepler* data. Among these, we identified strong activity cycle peaks in their power spectrum of 25 sources, with eight showing evidence of multiple cycles. Some RS CVn candidates display rapid rotation rates, the underlying cause of which might be tidal locking owing to the presence of a close companion.

In Fig. 9 we show the relationship between the activity cycle period and the rotation period for the RS CVn candidates in Table B2. The markers' sizes correspond to the cycle amplitudes. The cycle amplitude is lower for G giants than for K giants. Additionally, we do not observe a clear difference in cycle period between giants and dwarfs. Unlike main-sequence stars, the cycle amplitude in these evolved binary candidates does not vary significantly with increasing rotation period. This suggests that the dependence of dynamo action on rotation period may be less pronounced in giants compared to main-sequence stars. Alternatively, the observed trend in some stars could also result from the dilution of cycle amplitude owing to an increased number of faculae during the transition from spot to faculae dominance. However, it is important to note that our sample lacks a wide range of rotation periods for RS CVn candidates, thus limiting our ability to constrain the relationship(s) among rotation, stellar parameters, and cycle periods. The RS CVn candidates are anticipated to display distinctive dynamo behaviour given their specific evolutionary stage and the presence of a companion. For this work, we have restricted our sample of RS CVn variables to the available *Kepler* data. However, in the future, using ASAS-SN and ZTF data may provide an opportunity to detect activity cycles in larger samples of RS CVn variables, which will provide better observational constraints on their dynamo behaviour.

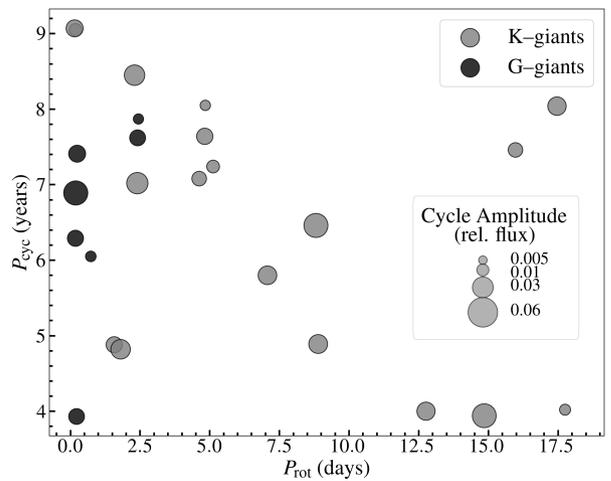

**Figure 9.** Rotation period ($P_{\rm rot}$) versus activity cycle periods ($P_{{\rm cyc},g}$) of G giants (black) and K giants (grey) from Table B2. The activity cycle amplitude (in relative flux) is indicated by the size of the data points, as specified in the legend.

## 5 DISCUSSION

### 5.1 How do activity cycles change?

Fast-rotating stars are believed to have a higher differential rotation (Benomar et al. 2018; Tokuno et al. 2023) and, hence, they are anticipated to be characterised by shorter dynamo cycles (Brown et al. 2007; Gastine et al. 2014; Brun et al. 2022). Recent research (Brandenburg et al. 2017) has demonstrated that fast-rotating stars display both shorter and longer cycles, suggesting the operation of two distinct types of dynamos. However, we did not observe a clear correlation between activity cycle period and stellar rotation, contrary to some expectations. On the other hand, our analysis revealed that several young, fast-rotating main-sequence stars display longer activity cycles ($\langle P_{{\rm cyc},g} \rangle \sim 6$ years), which was also observed by Suárez Mascareño et al. (2016). To check for shorter cycles in the fast-rotating sample, we investigated the peak in the power spectrum without binning the ZTF data, but we did not find any dominant shorter cycles. Even if such shorter cycles exist, we suspect that they may not be as dominant as the longer cycles in our sample.

To understand the dependence of activity cycles on stellar mass and rotation rate, we plotted rotation period versus temperature in Fig. 10. Figure 10 shows that young and fast-rotating main-sequence stars exhibit longer activity cycles ($P_{{\rm cyc},1}$), similar to our Sun. Moreover, there appears to be no strong correlation between cycle period and rotation for fast-rotating stars in our sample (refer to Fig. 10), suggesting that factors other than stellar rotation may also contribute to the cycle duration for stars in the fast-rotating regime. We observe that K-type stars exhibit longer cycle periods than G-type stars, likely owing to the presence of a thicker convective zone. Our results agree with Lehtinen et al. (2016) and Suárez Mascareño et al. (2016), who observed similar activity cycle periods in fast- and slowly rotating stars.

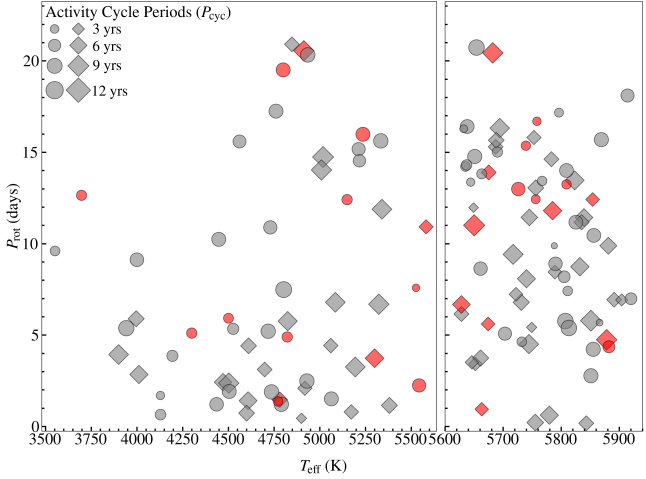

**Figure 10.** Rotation period versus effective temperature of cooler stars (left-hand panel) and Sun-like stars (right-hand panel). Grey data points represent stars with a single cycle ($P_{cyc,g,1}$), while red points indicate stars with multiple cycles (with $P_{cyc,g,1}$, plotted; $P_{cyc,g,2}$ can be found in Table B1). The sizes of the data points indicate the activity cycle periods ($P_{cyc,g,1}$) (in years); see the legends. The diamonds and circles represent cycle periods derived from fitting the *Kepler*–ASAS-SN–ZTF (first-set) and *Kepler*–ZTF (second-set) data sets, respectively, as listed in Table B1.

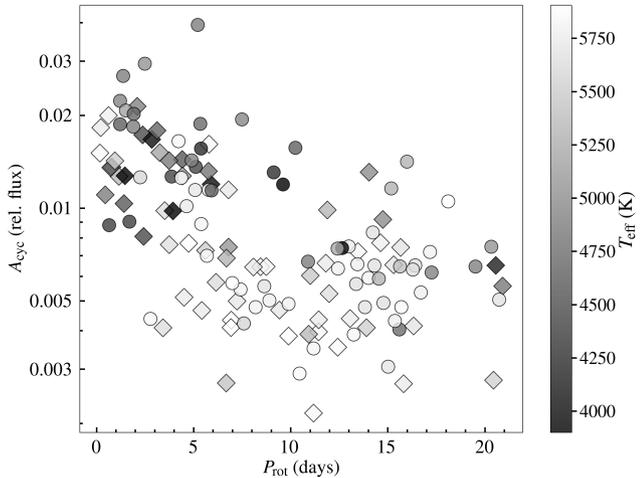

**Figure 11.** Activity Cycle Amplitude ($A_{cyc,g,1}$) (relative flux) against the rotation period ($P_{rot}$) of main-sequence stars listed in Table B1. The colour markers represent the effective temperature of stars. The diamonds and circles represent cycle periods derived from fitting the *Kepler*–ASAS-SN–ZTF (first-set) and *Kepler*–ZTF (second-set) data sets, respectively, as listed in Table B1.

Moreover, in Fig. 11 the activity-cycle amplitude and rotation seem to exhibit strong negative trends up to $P_{rot} \sim 10$ days, followed by a broad distribution (Saar & Brandenburg 2002). The cycle amplitude is lower for Sun-like stars, which suggests that this is caused by their thinner convection-zone depths compared with K-type stars. In addition, our sample does not indicate a good correlation between cycle period and cycle amplitude either, contrary to the conclusions of Irving et al. (2023) whose study included a significant fraction of slowly rotating stars. We do observe high-cycle amplitude for stars with $P_{rot} < 5$–10 days. However, when considering the fractional cycle amplitude, defined as the ratio of cycle amplitude to photospheric activity ($R_{per}$, taken from McQuillan et al. (2014)), we find no clear correlation with rotation period. This is consistent with the findings of Saar & Brandenburg (2002), who reported a similar lack of correlation in the fractional variation of chromospheric activity. Instead, we observe a very weak increase in fractional amplitude with rotation. Extending this study to more slowly rotating stars will help reveal a clearer dependence of cycle amplitude on both rotation and cycle periods.

The observed correlation of cycle amplitude with rotation period (see Fig. 11) agrees well with the trends observed in chromospheric activity indices for G and K-type stars by Chahal et al. (2023). The longer activity cycles in fast-rotating stars, along with their high cycle amplitudes, support the explanation of core–envelope coupling (Curtis et al. 2019; Spada & Lanzafame 2020; Curtis et al. 2020). The core–envelope coupling hypothesis outlined in Chahal et al. (2023) highlights that the convective envelope in K-type stars spins down to $P_{rot} \sim 2$–7 days, while the radiative core continues to rotate faster. The two zones start exchanging angular momentum around $P_{rot} \sim 15$–20 days, which leads to a decrease in shear/differential rotation and, hence, the observed steep decrease in activity and the origin of the period gap—a dearth of sources observed in the period–colour diagram in the *Kepler* (McQuillan et al. 2014), K2 (Reinhold & Hekker 2020; Gordon et al. 2021) and ZTF (Lu et al. 2022; Chahal et al. 2023) data sets.

We have observed large cycle amplitudes and longer cycles for K-type stars with $P_{rot} < 5$–10 days (see Fig. 11). We suggest that the increasing shear is caused by an increase in differential rotation between the two zones, since the convective envelope spins down while the radiative core is still rotating faster. Consequently, this gives rise to large cycle amplitudes. Conversely, for stars with $P_{rot} \sim 15$–20 days, we observed a $\sim 70$ per cent reduction in the cycle amplitudes (see Fig. 11). This reduction aligns with our hypothesis of reduced differential rotation owing to core–envelope coupling, leading to a steep decrease in activity for K–M type stars (see Chahal et al. 2023, their fig. 9). We did not observe a similar change in amplitude for G-type stars, consistent with Chahal et al. (2023). However, note that our sample is predominantly composed of fast-rotating stars. Therefore, expanding our investigation to include activity cycles of a larger sample of slowly rotating stars with periods exceeding 15–20 days could offer valuable insights into and further constrain the core–envelope coupling process.

In addition, our analysis shows that a significant fraction of fast-rotating stars in our sample exhibit longer activity cycles ($\langle P_{cyc,g} \rangle \sim 6$ years; see Fig. 10). Particularly, if a star exhibits multiple cycles, the shorter cycles are $\geq 3$–4 years in some cases. This indicates that a surface shear-layer dynamo might not be dominant in these stars, as proposed by Brandenburg et al. (2017). Instead, we suggest that a deep-seated convective dynamo likely drives the magnetic activity cycles, even in very fast-rotating stars. This is consistent with their dynamos being of a turbulent $\alpha^2$ type, where the dynamo actions originate in the convection zone, as suggested by Kitchatinov & Rüdiger (1999). Recently, Irving et al. (2023) also proposed that fast-rotating stars are likely governed by an $\alpha^2$- or $\alpha^2\Omega$-type dynamo mechanism. However, we also acknowledge that there maybe other dynamo models which can explain the occurrence of $\langle P_{cyc,g} \rangle \sim 6$ years in fast-rotating stars.

## 5.2 Investigation of the proposed active–inactive branches

To further investigate the relationships among and between cycle period, rotation, Rossby number and chromospheric activity indices, we explore the behaviour of these parameters in the following subsections.

### 5.2.1 Rotation – Cycle Period *versus* Chromospheric Activity

To investigate the relationship between stellar activity cycles and the dynamo process, different kinds of samples and plots have shown the existence of two branches, an active and an inactive branch, corresponding to shorter and longer activity cycles, respectively (Böhm-Vitense 2007). Brandenburg et al. (1998) and Böhm-Vitense (2007) found two branches with positive slopes in the $P_{cyc}/P_{rot}$ versus $R'_{HK}$ (instead of $\tau_{conv}/P_{rot}$) plot; $\langle R'_{HK} \rangle$ and $\tau_{conv}/P_{rot}$ are proportional to each other. In recent work, Brandenburg et al. (2017) found two branches in a similar diagram – active and inactive regions. Additionally, the authors observed multiple cycles in young stars. However, Boro Saikia et al. (2018) and Olspert et al. (2018) have argued that this classification into active and inactive branches might not be a true representation of the complex relationship between cycle periods and stellar rotation.

In addition to the $P_{cyc}$ versus $P_{rot}$ plot in Fig. 6, we have also plotted $P_{cyc}/P_{rot}$ versus $R'_{H\alpha}$ in Fig. 12, which is similar to fig. 5 of Brandenburg et al. (2017). For a plane-wave $\alpha\Omega$ dynamo, Brandenburg et al. (2017) suggested that the graph of $\omega_{cyc}/\Omega = P_{cyc}/P_{rot}$ versus $\langle R'_{HK} \rangle$ supports a theoretical interpretation in terms of the mean-field dynamo theory (Brandenburg et al. 2017). However, instead of $\langle R'_{HK} \rangle$, we derived $R'_{H\alpha}$ indices from the single-epoch LAMOST spectral observations, which might have introduced some scatter. The SNR is low for the Ca II H and K emission lines in comparison with the H$\alpha$ line in the low-resolution LAMOST spectra ($R = 1800$ at 5500 Å). In Fig. 12 we have plotted cycles classified as 'good' in Table B1; the shaded pink regions are the active (bottom) and inactive (top) regions observed by Brandenburg et al. (2017). We found a significant sample of stars lying well below the magnetically active region in Fig. 12, primarily because of their shorter rotation periods and the absence of a clear correlation between cycle and rotation periods in our sample. Moreover, approximately one-third of our sample is found within the intermediate region between both the regions, commonly referred to as the transition region, where our Sun lies as well. The sample of stars lying in the transition region is further discussed in Section 5.2.2. Additionally, we have included sources with multiple cycles, connected by dashed lines in Fig. 12. We note the presence of these multiple cycles below, above, as well as within both the magnetically active and inactive regions. Several sources even exhibit multiple cycles similar to solar-like Hale–Schwabe cycles ($P_{cyc,Hale} \approx 2P_{cyc,Schwabe}$), which is discussed in Section 5.3.

Furthermore, we have depicted the evolutionary trajectory of stars across $P_{cyc}/P_{rot}$ versus $R'_{H\alpha}$ space in Fig. 12. Recent open clusters-based data-driven gyrochronological frameworks (Bouma et al. 2023; Van-Lane et al. 2024) offer comparatively better age estimation. We inferred gyrochronological ages for our sample by employing an interpolation-based gyrochronology framework, *gyro-interp* (Bouma et al. 2023). The uncertainty in the ages for very fast-rotating stars exceeds 20–30 per cent. Figure 12 exhibits a grey-shaded distribution of ages, revealing the evolution of stars from below the magnetically-active towards the inactive region. Brandenburg et al. (1998) and Böhm-Vitense (2007) suggested that the activity cycles of stars situated on the active branch may migrate towards the inactive branch. However, we hypothesize that stars transition from the magnetically active to the quiescent region, which does not necessarily correspond to migration between the active towards inactive branch. This contrasts with the evolutionary tracks adopted by Brandenburg et al. (2017) in a similar diagram.

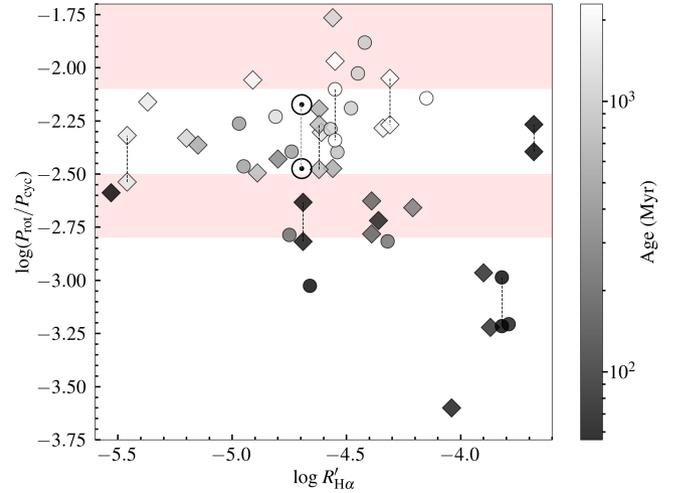

**Figure 12.** $\log(\omega_{cyc}/\Omega) = \log(P_{rot}/P_{cyc,g,1})$ versus chromospheric activity indices ($\log R'_{H\alpha}$) plot, similar to Brandenburg et al. (2017) for G and K-type stars. The diamonds and circles represent cycle periods derived from fitting the *Kepler*–ASAS-SN–ZTF (first-set) and *Kepler*–ZTF (second-set) data sets, respectively, from Table B1. Sources with multiple cycles are connected by dashed lines. The pink shaded regions correspond to the approximate active (bottom) and inactive (top) regions from Brandenburg et al. (2017). The colour markers represent gyrochronological ages derived using *gyro-interp* (Bouma et al. 2023), as per the grey-scale provided. The Sun is shown as ⊙, with its Hale–Schwabe cycles ($P_{cyc,Hale} \approx 2P_{cyc,Schwabe}$).

### 5.2.2 Stars in the Transition Region: *Cycle versus Rotation Period*

We found a significant fraction of stars in the intermediate region between the two branches in Fig. 6 & 12. We note that our target sample lacks a significant fraction of K-type stars with $P_{rot} > 10$ days, since these stars were sourced from our BY Dra catalogue of fast-rotating stars (Chahal et al. 2022). However, our sample does contain a substantial proportion of Sun-like stars with $P_{rot} > 10$ days taken from Montet et al. (2017). Sun-like stars are defined based on their effective temperatures within 150 K of the Sun. Specifically, our target sample comprises 57 per cent (79/138) Sun-like stars. To understand the dependence on mass for stars in the intermediate region, we plot their cycle period versus rotation period in Fig. 13. In Fig. 13, we found that 68 per cent (30/44) of the stars in the intermediate region are Sun-like stars. The activity-cycle amplitude for stars in this region is low in comparison to stars lying above the active branch. However, the ratio of cycle amplitude to mean photospheric activity remains consistent across these regions.

Previous studies by Böhm-Vitense (2007), Brandenburg et al. (2017), Boro Saikia et al. (2018) and Olspert et al. (2018) did not detect chromospheric cycles in a large sample of Sun-like stars. These stars exhibit emission within the Ca II H and K absorption lines owing to magnetic activity, making the detection of cyclic variations in Sun-like stars reliant on high-resolution, long-term spectral observations. We observe photometric variations of 0.3–0.8 per cent thanks to the precise and low-cadence photometry from *Kepler*, as well as

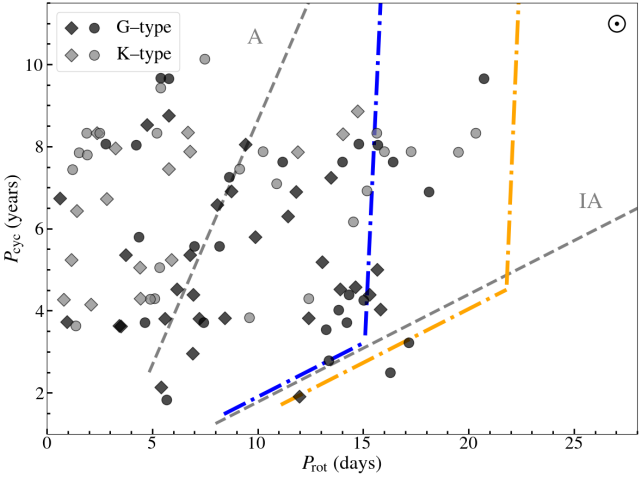

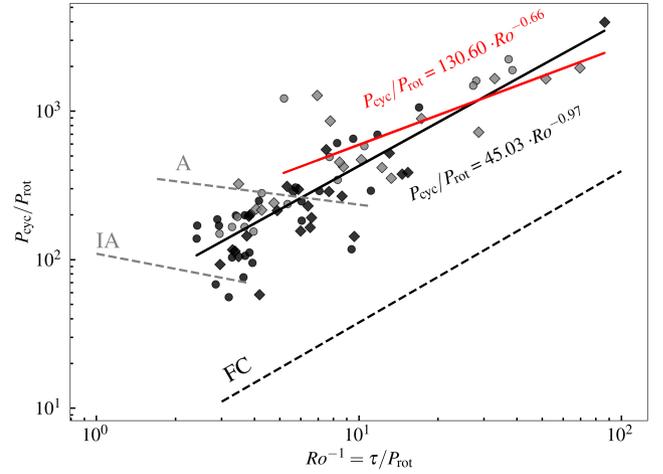

**Figure 13.** Activity-cycle period ($P_{cyc,g,1}$) as a function of rotation period ($P_{rot}$) for stars in Table B1. The diamonds and circles represent cycle periods derived from fitting, respectively, the *Kepler*–ASAS-SN–ZTF (first-set) and *Kepler*–ZTF (second-set) data sets, which are listed in Table B1. The black and grey data points correspond to G- and K-type main-sequence stars, respectively. The grey dashed lines show the active (A) and inactive (IA) branches according to Böhm-Vitense (2007). The blue and yellow dashed-dotted lines refer to the trajectories for F- and G-type stars, respectively, taken from Metcalfe & van Saders (2017). The Sun is shown as ⊙.

**Figure 14.** $P_{cyc,g,1}/P_{rot}$ versus the inverse of the Rossby number for G-type (black) and K-type (grey) stars from Table B1. The diamonds and circles represent cycle periods derived from fitting the *Kepler*–ASAS-SN–ZTF (first-set) and *Kepler*–ZTF (second-set) data sets, respectively. The grey dashed lines show the active (A) and inactive (IA) branches delineated by Irving et al. (2023). The black dashed line indicates the power-law fit for fully convective M dwarfs (FC; Irving et al. 2023). The black solid line represents the power-law fit to our sample. Additionally, the red solid line depicts the power-law fit for stars lying above the active branch, with their corresponding relation displayed.

ZTF. This has enabled the detection of activity cycles in several Sun-like stars, marking the highest number of such stars identified in the transition region where our Sun is located. Finding several young Sun-like stars in the transition region evidence that our Sun is likely not an exception and that the solar dynamo might not be in a transitional dynamo phase. We propose that using precise photometry, such as that provided by *Kepler* and TESS, can enable the detection of activity cycles in a larger sample of Sun-like stars, thereby filling the transition region with solar cycle analogues. This can provide a more clear understanding that our Sun is likely in a common evolutionary phase of the stellar dynamo, contrary to as suggested by Metcalfe et al. (2016).

We found that all Sun-like stars in the intermediate region fall well below the critical Rossby number limit ($Ro \sim 2$) defined by Metcalfe & van Saders (2017). Metcalfe (2017) suggested that as G-type stars spindown with the age their cycle periods evolve along two sequences until they reach a critical Rossby number ($Ro \sim 2$), where such stars begin their magnetic transition (see the blue and yellow evolutionary tracks for F- and G-type stars, respectively, in Fig. 13). However, we found most Sun-like (black) and K-type (grey) stars in the transition region at lower Rossby numbers than $Ro \sim 2$ (as suggested by Metcalfe & van Saders 2017), with the majority of these stars having Rossby number below 0.5. Detecting a substantial sample of Sun-like stars in the transition region well below the critical Rossby number suggests that cycle periods do not evolve at least along the active branch as stars spindown. However, our target sample cannot confirm whether Sun-like stars begin their magnetic transition at the critical Rossby number ($Ro \sim 2$), as proposed by Metcalfe & van Saders (2017); Jeffers et al. (2023). We suspect that younger stars may evolve downwards towards the intermediate region in Fig. 13, while older stars already on the inactive branch might continue evolving along it or evolve towards the intermediate region, until they reach the transition-off stage. Additionally, the stars might also evolve across $P_{cyc}/P_{rot}$ versus $R'_{H\alpha}$ space from the magnetically active to the inactive region (see Fig. 12 and subsection 5.2.1 for details). It is important to note that our target sample primarily consisted of fast-rotating stars (with $P_{rot} < 20$ days), further detecting activity cycles in slow-rotating Sun-like stars can provide better constraints on the evolutionary tracks of activity cycles, particularly in the context of our Sun.

*5.2.3* **Cycle – Rotation Period** *versus* **Rossby Number**

Figure 6 illustrates the relationship between the cycle period and rotation rate, revealing a notable dispersion without a strong correlation. The observed scatter may be attributed to a lack of consideration for convective zones, which are the key ingredient for stellar cycles. To address this limitation, we computed the Rossby number using the empirical relation of Corsaro et al. (2021), which employs *Gaia* colours. Irving et al. (2023) recently studied dynamo behaviour by examining the ratio of the cycle-to-rotation period alongside the inverse of the Rossby number. In Fig. 14, we present the cycle period divided by the rotation period plotted against the inverse of the Rossby number, focusing on cycles exhibiting prominent single peaks and classified as 'good' in Table B1.

We have fitted the data with power-law fits for all samples of stars (see the black solid line in Fig. 14), which follows the relation

$$\frac{P_{cyc}}{P_{rot}} = 45.03 \pm 1.53 \times Ro^{-0.97 \pm 0.02}, \quad (1)$$

suggesting that $P_{cyc} \propto \tau_{conv}$ for all stars, implying that this is merely a function of convection zone properties. We have also plotted the active (A) and inactive (IA) branches of Saar & Brandenburg (1999). The power-law index we obtained, $-0.97 \pm 0.02$, matches closely the Irving et al. (2023) indices of $-1.06$ and $-1.02$ observed for FGK and M dwarfs, respectively. However, our best-fitting scale factor is 0.71 times their scale factor for FGK stars.

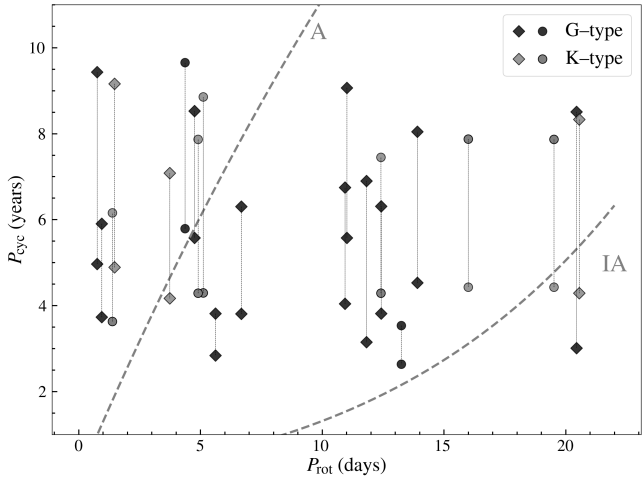

**Figure 15.** Relationship between cycle periods ($P_{\text{cyc},g,1}$ & $P_{\text{cyc},g,2}$) and rotation periods ($P_{\text{rot}}$) for G-type (black) and K-type (grey) stars exhibiting multiple cycles, from Table B1. The diamonds and circles represent cycle periods derived from fitting the *Kepler*–ASAS-SN–ZTF (first-set) and *Kepler*–ZTF (second-set) data sets, respectively. The grey dashed lines representing active (A) and inactive (IA) branches were taken from do Nascimento et al. (2023). Multiple cycles observed for the same star are connected by vertical dotted lines.

Our data set complements that of Irving et al. (2023), particularly because we analyse a substantial number of stars exhibiting very fast rotation. Consequently, we fitted a power law to the stars positioned above the active branch (see the red solid line in Fig. 14), resulting in the relation,

$$\frac{P_{\text{cyc}}}{P_{\text{rot}}} = 130.60 \pm 2.99 \times Ro^{-0.66 \pm 0.07}. \qquad (2)$$

The best-fitting power-law index remains consistent with Irving et al. (2023). However, the scale factor for stars positioned above the active branch is 0.83 times that of Irving et al. (2023). This suggests that our sample occupies the gap identified by Irving et al. (2023) between FGK and M dwarfs (the fit to the fully convective M dwarfs, 'FC', is shown in Fig. 14; cf. Irving et al. 2023). This is primarily so, because our sample includes cooler K-type stars, which were not covered by Irving et al. (2023). Our results corroborate the discussions regarding the dynamo process in fast-rotating stars (Irving et al. 2023). However, we observe a sample of stars displaying multiple cycles (refer to Fig. 7), which were not observed by Irving et al. (2023).

### 5.3 Do multiple cycles correspond to two branches?

The Sun exhibits various cycles, with the 11-year cycle being the most prominent (Krivova et al. 2006). Additionally, long-term variations in sunspot numbers reveal an 80-year cycle known as the Gleissberg cycle, along with periodicities of 51.34 years, 8.83 years (known as the short-term or sub-decadal cycle) and 3.77 years (the intermediate-term or quasi-biennial cycle; Deng et al. 2016). Boro Saikia et al. (2018) detected multiple cycles through long-term measurements of chromospheric activity indices. A recent study (do Nascimento et al. 2023) found that 18 Sco, a young Sun-like star, exhibits paired magnetic and activity cycles similar to those of the Sun, with periods that appear to be twinned, $P_{\text{cyc,Hale}} \approx 2 P_{\text{cyc,Schwabe}}$.

We have identified dual cycle peaks in the power spectra of a sample comprising 30 G and K-type stars, listed in Table B1. Interestingly, in several stars, we observed that the period of the second peak is approximately twice or half that of the strongest peak. This could be attributed to the presence of Hale–Schwabe twin cycles ($P_{\text{cyc,Hale}} \approx 2 P_{\text{cyc,Schwabe}}$), similar to those observed in the Sun and 18 Sco. However, they could also be harmonics of each other. This is further discussed in Appendix B.

We have additionally investigated the behaviour of these dual peaks by plotting them in Fig. 15, presuming they represent distinct activity cycles. Few of these dual cycles exhibit resemblance to solar-like twin cycles ($P_{\text{cyc,Hale}} \approx 2 P_{\text{cyc,Schwabe}}$); see Fig. 15. Our sample is mainly confined to stars with $P_{\text{rot}} < 15$ days, and we have not considered $P_{\text{cyc}} > 10$ years. Contrary to previous studies (e.g., Saar & Brandenburg 1999; Brandenburg et al. 2017; do Nascimento et al. 2023), we did not observe dual cycles aligned with the active and inactive branches (see Fig. 15).

## 6 CONCLUSIONS

We have presented a sample of 138 G–K-type stars that exhibit photometric activity cycles, comprised of first set with 70 sources with cycles derived from combined *Kepler*–ASAS-SN–ZTF data and second set of 68 sources with cycles derived from *Kepler*–ZTF data. By combining light curves from *Kepler* FFIs and ASAS-SN with the ZTF data, we extended the observational baseline to 14 years, thus allowing us to detect longer activity cycles. To remove the rotational modulation effect from the light curves, we applied binning to the photometric data using $n \times P_{\text{rot}}$ (where $n = 1, 2, 3, ...$). The average FAP of the cycle periods in our data set is $\sim 10^{-4}$. Our analysis has successfully identified both single and multiple cycles in fast-rotating G–K-type stars. Furthermore, we have determined cycle periods for a subset of 25 RS CVn candidates.

We found that several very fast-rotating stars exhibit longer activity cycle periods ($\langle P_{\text{cyc}}\rangle \sim 6$ years), hinting at the prevalence of a deeper convective dynamo mechanism. Interestingly, we have not found a strong correlation between the activity-cycle period and the rotation period for stars along the fast-rotating sequence. However, we observe a strong negative correlation between cycle amplitude and both rotation rate and stellar mass. In contrast, the fractional cycle amplitude shows no clear correlation with rotation period. Additionally, based on RMS variations, we have identified nine fast-rotating K-type stars that are faculae-dominated, particularly by tracking spot/faculae evolution in *Kepler* RMS data. While we have discussed the limitations of using RMS variations to identify spot/faculae dominance, we aim to further investigate this by incorporating chromospheric variability in future. Our observations align with literature regarding the relationship between cycle period, rotation and Rossby number (Suárez Mascareño et al. 2016; Irving et al. 2023).

We find that fast-rotating stars do not follow any (linear) trend between cycle period and rotation, particularly with respect to the active branch. Intriguingly, we have found several stars positioned above the active branch, suggesting its reality might not be as certain as previously thought, which is consistent with the observations of Boro Saikia et al. (2018). Moreover, we found 44 Sun-like and K-type stars lying within the transitional zone between the active and inactive branches, where our Sun resides. Notably, 30 of these stars are young Sun-like stars that fall well below the critical Rossby number limit ($Ro \sim 2$) defined by Metcalfe & van Saders (2017), contradicting their evolutionary tracks. These young Sun-like stars exhibit photometric variation of 0.3-0.8 per cent, detectable due to high precision photometry by *Kepler* and ZTF. Additionaly, we found cycle periods lying close to that of the Sun in the $P_{\text{cyc}}/P_{\text{rot}}$ versus $R'_{\text{HK}}$ plot between the magnetically active and quiescent regions.

This strongly implies that the solar dynamo is likely a common type of stellar dynamo. These results therefore suggest that the distinction between the two branches may not be as definitive as previously thought, particularly in relation to the active branch for the fast-rotating stars.

Considering the limitations of our data set and the considerable uncertainties in gyrochronological age estimates, Fig. 12 highlights that the stars in our sample might migrate from the magnetically active towards the inactive region (see subsection 5.2.1 for details). In Cycle-rotation plot (see Fig. 13), we suspect younger stars may descend toward the intermediate region, while older stars on the inactive branch likely evolve along it or towards the intermediate region until reaching the transition-off stage. Although our sample contains fast-rotating stars, adding a sample of slowly rotating stars can further constrain the existence of these two branches and the activity cycle evolutionary tracks. We also detected multiple cycles that display some resemblance to solar-like twin cycles, known as Hale–Schwabe cycles ($P_{\rm cyc,Hale} \approx 2P_{\rm cyc,Schwabe}$). However, further investigation of these dual cycles is required to confirm whether they indeed represent two activity cycles or if they are harmonics of one another.

Our next goal is to expand our research by deriving photometric activity cycles for a wider range of stars, including those at different evolutionary stages, with a particular focus on RS CVn candidates. Additionally, we plan to analyze the activity cycles of Sun-like stars and fully convective M dwarfs, which will provide valuable insights into solar–stellar dynamo mechanisms.

## ACKNOWLEDGEMENTS

D. C. acknowledges funding support from the International Macquarie Research Excellence Scheme (iMQRES). D.K acknowledges funding support from the ARC Discovery Project DP240101150. This research was supported in part by the Australian Research Council Centre of Excellence for All Sky Astrophysics in 3 Dimensions (ASTRO 3D), through project number CE170100013. We are indebted to the referee, Steven H. Saar, for providing invaluable suggestions and significantly contributing to improvements to our manuscript.

## DATA AVAILABILITY

The ZTF DR20, *Kepler* and ASAS-SN data analysed in this paper are publicly available from the ZTF, *Kepler* and ASAS-SN Archives, respectively. The data forming the basis for this article are available online at https://doi.org/10.5281/zenodo.15336130. Additionally, the data are also available in the article and in its online supplementary material.

## REFERENCES

Babcock H. W., 1961, ApJ, 133, 572
Bailer-Jones C. A. L., Rybizki J., Fouesneau M., Demleitner M., Andrae R., 2021, AJ, 161, 147
Baliunas S. L., et al., 1995, ApJ, 438, 269
Baliunas S. L., Nesme-Ribes E., Sokoloff D., Soon W. H., 1996, ApJ, 460, 848
Bellm E. C., et al., 2018, PASP, 131, 018002
Benomar O., et al., 2018, Science, 361, 1231
Berdyugina S. V., Järvinen S. P., 2005, Astronomische Nachrichten, 326, 283
Berdyugina S. V., Tuominen I., 1998, A&A, 336, L25
Böhm-Vitense E., 2007, ApJ, 657, 486
Boro Saikia S., et al., 2018, A&A, 616, A108
Bouma L. G., Palumbo E. K., Hillenbrand L. A., 2023, ApJ, 947, L3
Brandenburg A., Saar S. H., Turpin C. R., 1998, ApJ, 498, L51
Brandenburg A., Mathur S., Metcalfe T. S., 2017, ApJ, 845, 79
Brown B. P., Browning M. K., Brun A. S., Miesch M. S., Nelson N. J., Toomre J., 2007, in AIP Conference Proceedings. pp 271–278
Brun A. S., Strugarek A., Noraz Q., Perri B., Varela J., Augustson K., Charbonneau P., Toomre J., 2022, ApJ, 926, 21
Chahal D., de Grijs R., Kamath D., Chen X., 2022, MNRAS, 514, 4932
Chahal D., Kamath D., de Grijs R., Ventura P., Chen X., 2023, MNRAS, 525, 4026
Charbonneau P., 2010, Living Reviews in Solar Physics, 7, 3
Chen X., Wang S., Deng L., de Grijs R., Yang M., Tian H., 2020, ApJS, 249, 18
Coffaro M., et al., 2020, A&A, 636, A49
Corsaro E., Bonanno A., Mathur S., García R. A., Santos A. R. G., Breton S. N., Khalatyan A., 2021, A&A, 652, L2
Cram L. E., Giampapa M. S., 1987, ApJ, 323, 316
Cui X.-Q., et al., 2012, Research in Astronomy & Astrophysics, 12, 1197
Curtis J. L., Agüeros M. A., Douglas S. T., Meibom S., 2019, ApJ, 879, 49
Curtis J. L., et al., 2020, ApJ, 904, 140
Das P. B., et al., 2025, MNRAS, 538, 605
Deng L. H., Xiang Y. Y., Qu Z. N., An J. M., 2016, AJ, 151, 70
Dikpati M., Gilman P. A., 2006, ApJ, 649, 498
Distefano E., Lanzafame A. C., Lanza A. F., Messina S., Spada F., 2017, å, 606, A58
Feinstein A. D., Montet B. T., Johnson M. C., Bean J. L., David T. J., Gully-Santiago M. A., Livingston J. H., Luger R., 2021, AJ, 162, 213
Frasca A., et al., 2016, A&A, 594, A39
Fröhlich C., Lean J., 2004, A&ARv, 12, 273
Gaia Collaboration et al., 2016, A&A, 595, A1
Gaia Collaboration et al., 2023a, A&A, 674, A1
Gaia Collaboration et al., 2023b, A&A, 674, A34
Gastine T., Yadav R. K., Morin J., Reiners A., Wicht J., 2014, MNRAS, 438, L76
Gilliland R. L., et al., 2011, ApJS, 197, 6
Gomes da Silva J., Figueira P., Santos N., Faria J., 2018, The Journal of Open Source Software, 3, 667
Gomes da Silva J., et al., 2021, A&A, 646, A77
Gomes da Silva J., Bensabat A., Monteiro T., Santos N. C., 2022, A&A, 668, A174
Gordon T. A., Davenport J. R. A., Angus R., Foreman-Mackey D., Agol E., Covey K. R., Agüeros M. A., Kipping D., 2021, ApJ, 913, 70
Graham M. J., et al., 2019, PASP, 131, 078001
Gully-Santiago M. A., et al., 2017, ApJ, 836, 200
Hart K., et al., 2023, arXiv e-prints
Hathaway D. H., 2010, Living Reviews in Solar Physics, 7, 1
Irving Z. A., Saar S. H., Wargelin B. J., do Nascimento J.-D., 2023, ApJ, 949, 51
Isaacson H., et al., 2024, ApJS, 274, 35
Jayasinghe T., et al., 2018, MNRAS, 477, 3145
Jayasinghe T., et al., 2020, MNRAS, 491, 13
Jayasinghe T., et al., 2021, MNRAS, 503, 200
Jeffers S. V., Kiefer R., Metcalfe T. S., 2023, Space Science Reviews, 219, 54
Kitchatinov L. L., Rüdiger G., 1999, A&A, 344, 911
Krivova N. A., Solanki S. K., Floyd L., 2006, A&A, 452, 631
Lehtinen J., Jetsu L., Hackman T., Kajatkari P., Henry G. W., 2016, A&A, 588, A38
Leighton R. B., 1964, ApJ, 140, 1547
Lockwood G. W., Skiff B. A., Henry G. W., Henry S., Radick R. R., Baliunas S. L., Donahue R. A., Soon W., 2007, ApJS, 171, 260
Lomb N. R., 1976, Astrophysics & Space Science, 39, 447
Lu Y. L., Curtis J. L., Angus R., David T. J., Hattori S., 2022, AJ, 164, 251
Masci F. J., et al., 2018, PASP, 131, 018003
Mathur S., Salabert D., García R. A., Ceillier T., 2014, Journal of Space Weather and Space Climate, 4, A15

McQuillan A., Mazeh T., Aigrain S., 2013, ApJ, 775, L11
McQuillan A., Mazeh T., Aigrain S., 2014, ApJS, 211, 24
Metcalfe T. S., van Saders J., 2017, Solar Physics, 292, 126
Metcalfe T. S., Egeland R., van Saders J., 2016, ApJ, 826, L2
Meunier N., Kretzschmar M., Gravet R., Mignon L., Delfosse X., 2022, A&A, 658, A57
Montet B. T., Simon J. D., 2016, ApJ, 830, L39
Montet B. T., Tovar G., Foreman-Mackey D., 2017, ApJ, 851, 116
Namekata K., et al., 2020, ApJ, 891, 103
Noyes R. W., Hartmann L. W., Baliunas S. L., Duncan D. K., Vaughan A. H., 1984, ApJ, 279, 763
Oláh K., et al., 2009, A&A, 501, 703
Oláh K., Kővári Z., Petrovay K., Soon W., Baliunas S., Kolláth Z., Vida K., 2016, A&A, 590, A133
Olspert N., Lehtinen J. J., Käpylä M. J., Pelt J., Grigorievskiy A., 2018, A&A, 619, A6
Parker E. N., 1955, ApJ, 122, 293
Radick R. R., Lockwood G. W., Henry G. W., Hall J. C., Pevtsov A. A., 2018, ApJ, 855, 75
Recio-Blanco A., et al., 2023, A&A, 674, A29
Reinhold T., Hekker S., 2020, A&A, 635, A43
Reinhold T., Cameron R. H., Gizon L., 2017, å, 603, A52
Reinhold T., Bell K. J., Kuszlewicz J., Hekker S., Shapiro A. I., 2019, å, 621, A21
Saar S. H., Brandenburg A., 1999, ApJ, 524, 295
Saar S. H., Brandenburg A., 2002, Astronomische Nachrichten, 323, 357
Salabert D., et al., 2016, A&A, 596, A31
Scargle J. D., 1982, ApJ, 263, 835
Shapiro A. I., Solanki S. K., Krivova N. A., Yeo K. L., Schmutz W. K., 2016, å, 589, A46
Shappee B. J., et al., 2014, ApJ, 788, 48
Solanki S. K., Inhester B., Schüssler M., 2006, Reports on Progress in Physics, 69, 563
Spada F., Lanzafame A. C., 2020, A&A, 636, A76
Stassun K. G., Torres G., 2021, ApJ, 907, L33
Suárez Mascareño A., Rebolo R., González Hernández J. I., 2016, A&A, 595, A12
Tokuno T., Suzuki T. K., Shoda M., 2023, MNRAS, 520, 418
Van-Lane P. R., et al., 2024, arXiv e-prints
Wilson O. C., 1968, ApJ, 153, 221
Wilson O. C., 1978, ApJ, 226, 379
Xu F., Gu S., Ioannidis P., 2021, MNRAS, 501, 1878
Yan L., et al., 2023, AGU Advances, 4
Zechmeister M., Kürster M., 2009, A&A, 496, 577
de Grijs R., Kamath D., 2021, Universe, 7, 440
do Nascimento J. D., et al., 2023, ApJ, 958, 57
van Saders J. L., Ceillier T., Metcalfe T. S., Silva Aguirre V., Pinsonneault M. H., García R. A., Mathur S., Davies G. R., 2016, Nature, 529, 181

## APPENDIX A: PHOTOMETRIC CYCLE FITS

In Fig. A1, we compared the cycle periods determined using the *Kepler* FFIs by Montet et al. (2017) with the one derived using the ZTF data. The cycle periods show a strong one-to-one correlation, except for one or two sources, which is discussed in subsection 2.3.

In Section 3, we showed a limited sample of sources fitted with single-cycle periods derived from their power spectra (see Fig. 3). Expanding upon this, here we present a more extensive selection of sources fitted with single-peak periods: see Fig. A2. We also explore sources exhibiting multiple cycles in their power spectra, as discussed in Section 3.2.1, and illustrate a sample of such sources alongside their power spectra in Fig. A3.

Note that only a subset of fits are presented here. However, all fits to our 138 main-sequence stars and 25 RS CVn candidates can be accessed online[1].

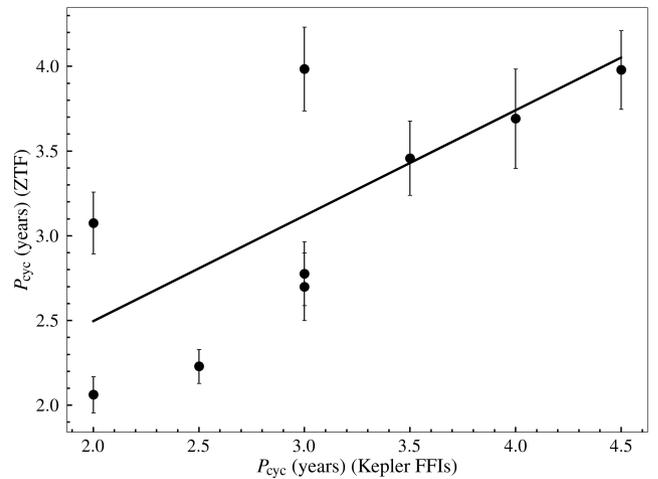

**Figure A1.** Comparison of photometric cycle periods derived by Montet et al. (2017) using the *Kepler* FFIs and those obtained from the ZTF data. The black line shows the one-to-one correlation between both cycle periods.

## APPENDIX B: METHODOLOGY FOR CONFIRMING MULTIPLE CYCLES

In Section 3.2.1 we discuss our observations of multiple strong peaks in the power spectrum. A sample of such sources is shown in Fig. A3. We detected 30 sources exhibiting such strong double peaks. The FAPs of both $P_{cyc,1}$ and $P_{cyc,2}$ are low: see Table B1. One possibility is that these stars exhibit two apparent activity cycles, which were also detected by Brandenburg et al. (2017) and do Nascimento et al. (2023). Alternatively, it is plausible that the dual peaks represent harmonics or aliased signals. To further check whether one of the two periods is the real period ($P_r$), whereas the other may be an aliased period ($P_a$), we show a comparison of both cycle periods in Fig. B1.

The relationship between the real ($P_r$) and aliased periods ($P_a$) predominantly adheres to three distinct equations: $1/P_a = |1/P_c \pm 1/P_r|$, where $P_c$ takes the values 0.5, 1 and 2, as indicated by the blue dotted line (when the relation follows '−') and the dashed line (when the relation follows '+') in Fig. B1, as suggested by Chen et al. (2020). If we further reduce the $P_c$ values to 0.125 and 0.0625, we find that our sample stars lie within $P_c \equiv [0.125–0.0625]$, shown as black and red lines, respectively, in Fig. B1. Additionally, we estimated the beat period and analyzed its relationship with the cycle periods, finding that nearly 80 per cent of the beat periods align with either twice or half of the cycle period. Notably, the Sun's solar cycle was approximately 8 years during the Maunder Minimum (1645–1715 AD; Yan et al. 2023). Considering this variation in solar cycle period suggests that our sample might also have presence of two distinct solar-like twin cycles (Hale–Schwabe cycles); see Section 5.3. Nevertheless, for our current analysis, we have opted for the period with the lowest FAP as the most probable (real) cycle period, as discussed in Section 3.

To further examine the behaviour of the observed multiple cycles, in Fig. B2, we show the sources exhibiting single and multiple cycles, from Table B1. Figure 6 contains only stars showing single prominent cycles. Considering that dual periods depict two apparent cycles, we observe a scattering of multiple cycles beyond the active branch and within the transition region between the two branches in Fig. B2, similar to those observed in Fig. 6. Notably, the dispersion within the transition region is more pronounced among the G-type stars, owing to the scarcity of significant numbers of K-type stars with periods > 10 days in our sample. However, we have identified a sample of stars lying above the active branch which exhibit multiple cycles (see Fig. 15).

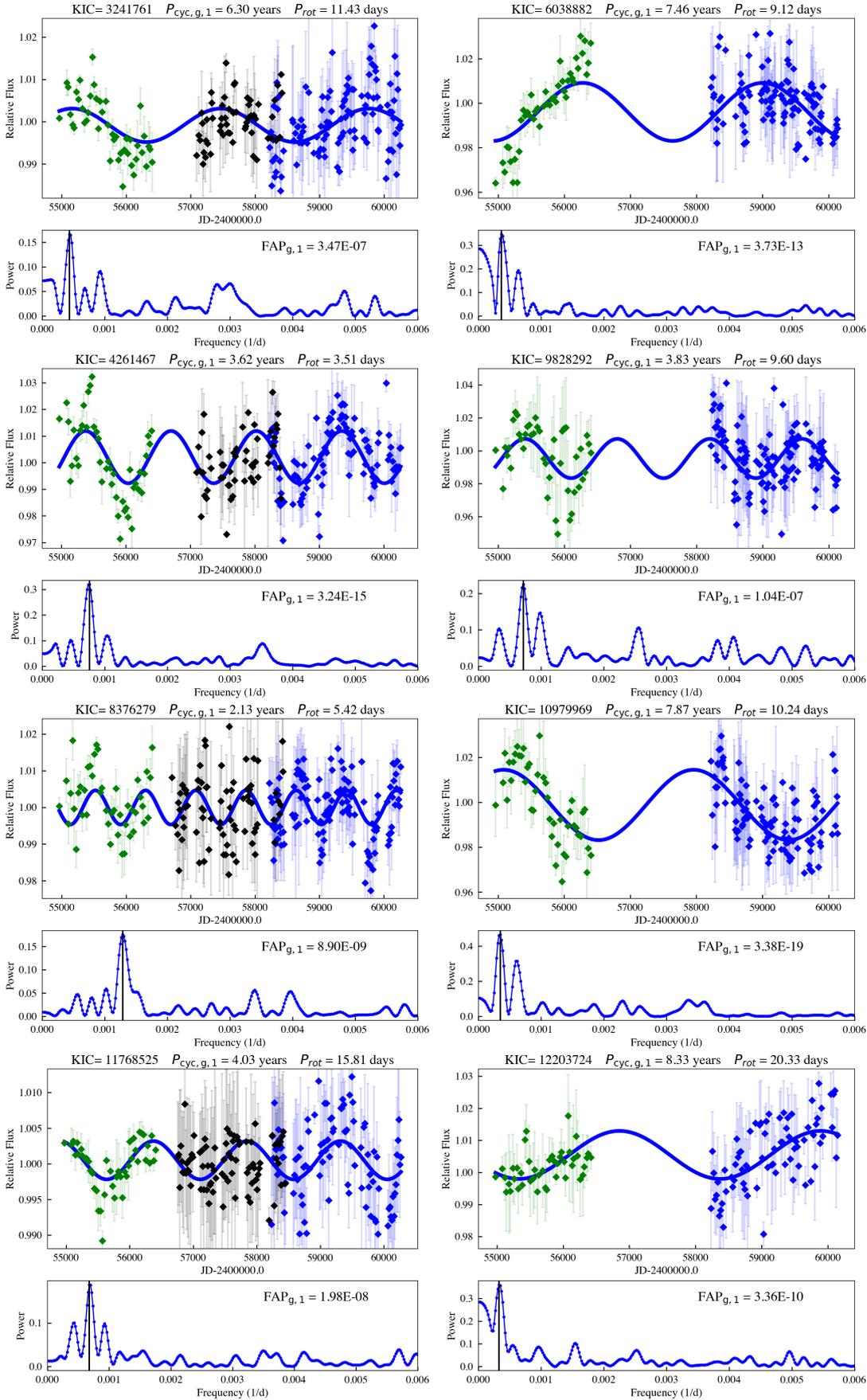

**Figure A2.** Examples of photometric cycle fitting to the combined *Kepler* (green)–ASAS-SN (black)–ZTF *g*-band (blue) (first-set) data in the left-hand panels and *Kepler* (green)–ZTF *g*-band (blue) (second-set) data in the right-hand panels. The error bars on the photometric fits represent the standard deviation within each bin. Below each photometric data panel we have included the power spectrum displaying the peak corresponding to the best-fitting period, along with the FAP of the cycle peak.

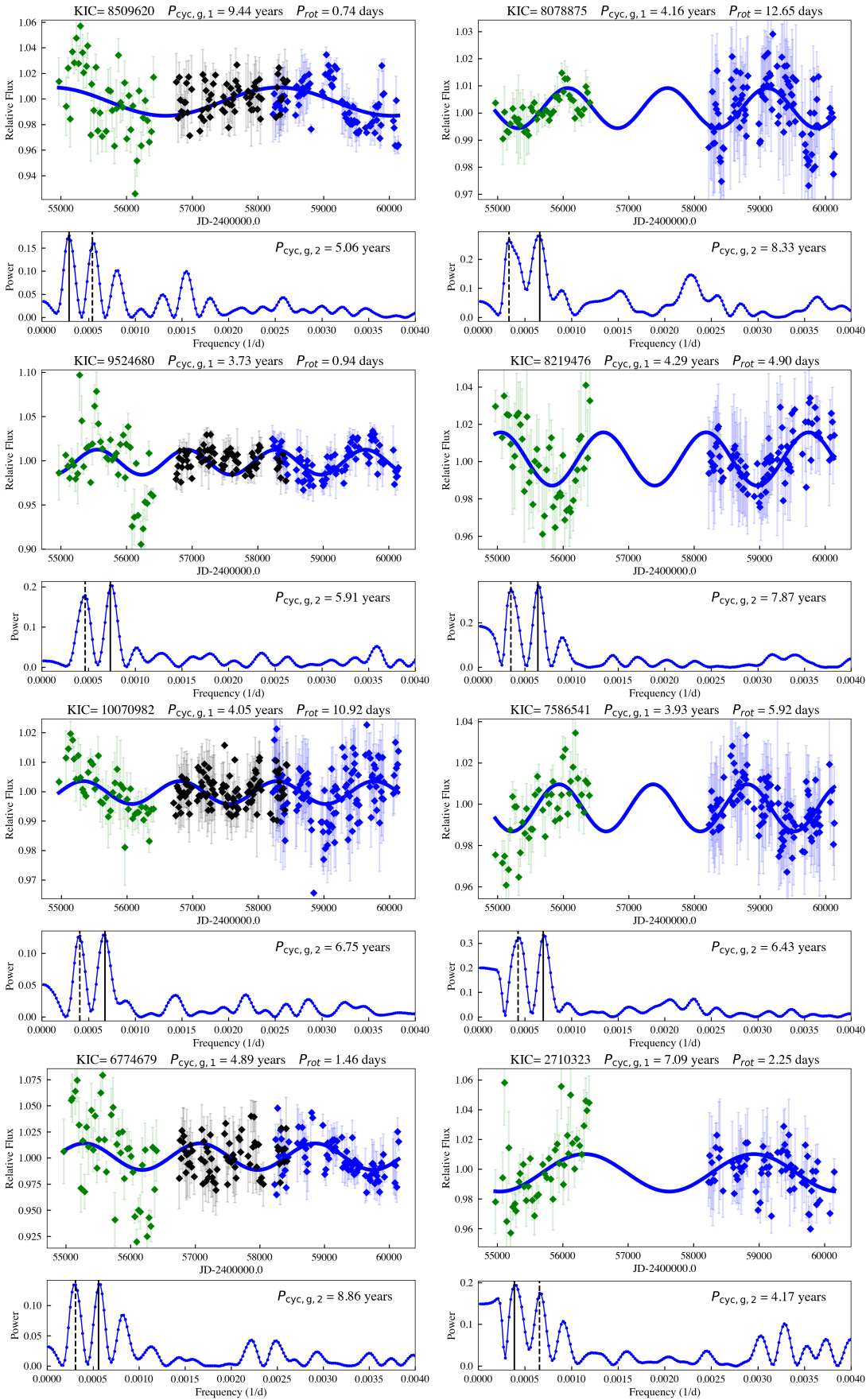

**Figure A3.** Examples of photometric cycle fitting to the combined *Kepler* (green)–ASAS-SN (black)–ZTF *g*-band (blue) (first-set) data in the left-hand panels and *Kepler* (green)–ZTF *g*-band (blue) (second-set) data in the right-hand panels, for eight sources that display multiple cycles. Below each photometric data panel we show the power spectrum displaying two strong peaks, $P_{\text{cyc},1}$ (solid line) and $P_{\text{cyc},2}$ (dashed line). The error bars on the photometric fits represent the standard deviation within each bin.

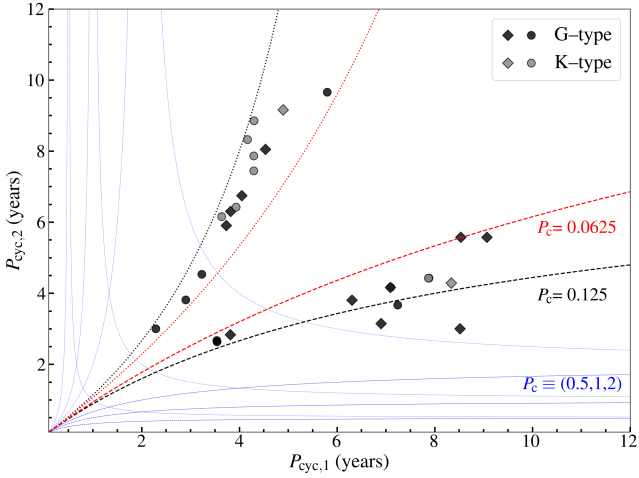

**Figure B1.** Comparison of activity-cycle periods corresponding to double peaks in the power spectrum, where the strongest peak with the lowest FAP is $P_{cyc,1}$ and the other strong peak is $P_{cyc,2}$. The plot follows the relationship between real ($P_r$) versus aliased ($P_a$) periods defined by the relation $1/P_a = |1/P_c \pm 1/P_r|$. The different colours and line styles in the plot are as follows: the dashed line indicates when the relation follows '+', the dotted line indicates when it follows '−', and the red, black and blue colours correspond to $P_c = 0.0625$, $0.125$ and $[0.5, 1, 2]$, respectively. The data fall between $P_c \equiv [0.0675 - 0.125]$. See the text for details.

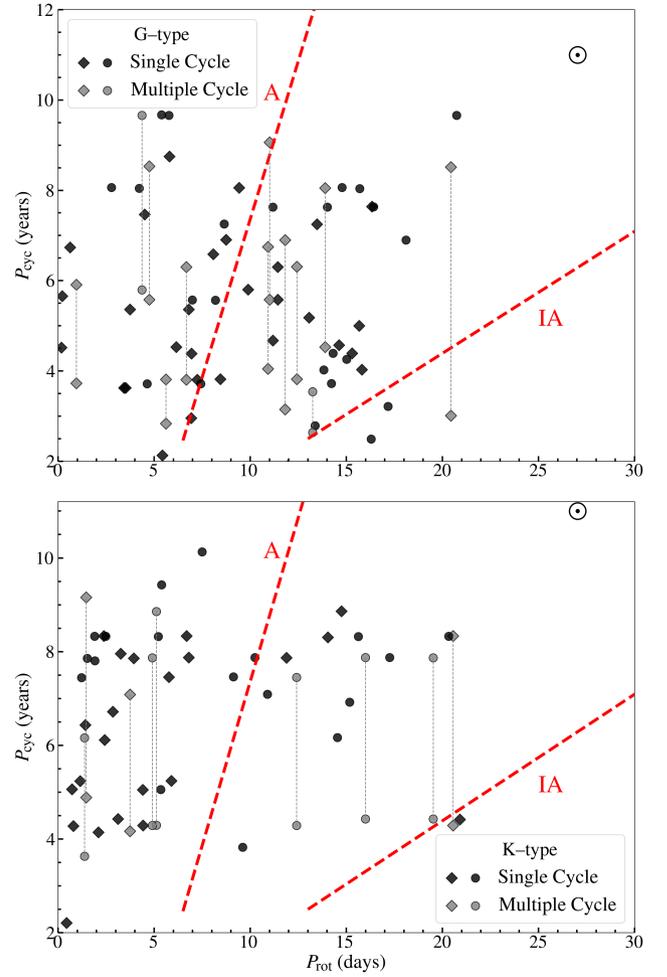

**Figure B2.** Activity-cycle period as a function of rotation period for (top) G dwarfs and (bottom) K dwarfs, from Table B1 (similar to Fig. 6). The black and grey symbols are the sources with single and multiple cycles, respectively, listed in Table B1. Stars with multiple cycles ($P_{cyc,1}$ and $P_{cyc,2}$) are connected by dashed lines. The red dashed lines show the active (A) and inactive (IA) branches according to Böhm-Vitense (2007). The Sun is shown as ⊙.

**Table B1.** Compiled table of 138 main-sequence stars whose activity cycles were derived from combined photometry for the first set using *Kepler*-FFIs–ASAS-SN–ZTF data (70 sources) and the second set using *Kepler*-FFIs–ZTF data (68 sources) with their cycle periods and best-fitting parameters (Cycle Amplitude, FAP, phase, etc.) for both ZTF *g* and *r* bands (see Fig. 2 for reference). Additionally, we also list activity cycles for sources with double peaks. Note that we have referred 5 sources as 'binary candidates' in the 'Fit Quality' column. Only a portion of the table is shown here. The full table is available online[1].

| KIC | ZTF OID$_g$ | RA (J2000) (°) | Dec (J2000) (°) | $P_{rot}$ (days) | $T_{eff}$ (K) | $P_{cyc,g,1}$ (years) | Error $P_{cyc,g,1}$ (years) | FAP$_{cyc,g,1}$ | $A_{cyc,g,1}$ (rel flux) | $\cos\phi_{cyc,g,1}$ | $P_{cyc,r,1}$ (years) | FAP$_{cyc,r,1}$ | ... | $\log R'_{H\alpha}$ | $P_{cyc,g,2}$ (years) | ... | Spot/Faculae | Fit Quality |
|---|---|---|---|---|---|---|---|---|---|---|---|---|---|---|---|---|---|---|
| 8078875 | 726113100002538 | 283.3180 | 43.9641 | 12.65 | 3697 | 4.16 | 0.14 | 2.69E–9 | 0.0074 | –0.92 | 4.43 | 1.17E–9 | ... | – | 8.33 | ... | Spot | Average |
| 1028018 | 727101400037244 | 291.4713 | 36.7998 | 0.62 | 5779 | 6.74 | 0.32 | 1.07E–12 | 0.0199 | 0.83 | 6.75 | 1.26E–12 | ... | –4.04 | – | ... | Faculae | Good |
| 6038882 | 727109300002815 | 290.6294 | 41.3396 | 9.12 | 4000 | 7.46 | 0.31 | 3.72E–13 | 0.0131 | 0.89 | 7.87 | 3.75E–9 | ... | – | – | ... | – | Good |
| 7433457 | 727114300007048 | 288.0918 | 43.0675 | 5.18 | 4943 | 10.12 | 0.53 | 3.81E–15 | 0.0137 | –0.92 | 9.44 | 1.4E–16 | ... | –4.14 | – | ... | – | Binary-candidate |
| 8894773 | 766102100039143 | 295.0384 | 45.1448 | 1.89 | 4736 | 8.32 | 0.44 | 4.9E–13 | 0.0184 | –0.98 | 7.83 | 2.8E–8 | ... | –3.79 | – | ... | Spot | Good |
| ... | ... | ... | ... | ... | ... | ... | ... | ... | ... | ... | ... | ... | ... | ... | ... | ... | ... | ... |

**Table B2.** Compiled table of 25 RSCVn candidates whose activity cycles were derived from combined photometry of *Kepler*-FFIs–ASAS-SN–ZTF (7 sources) and *Kepler*-FFIs–ZTF (18) with cycle periods and best-fitting parameters (Cycle Amplitude, FAP, phase, etc.) for both ZTF *g* and *r* bands. The full table is available online[1].

| KIC | ZTF OID$_g$ | RA (J2000) (°) | Dec (J2000) (°) | $P_{rot}$ (days) | $P_{cyc,g}$ (years) | Error $P_{cyc,g}$ (years) | FAP$_{cyc,g}$ | $A_{cyc,g}$ (rel. flux) | $\cos\phi_{cyc,g}$ | $P_{cyc,r}$ (years) | FAP$_{cyc,r}$ | ... | Single/Multiple Cycles | Spot/Faculae | Fit Quality |
|---|---|---|---|---|---|---|---|---|---|---|---|---|---|---|---|
| 7337049 | 726113300005664 | 281.6682 | 42.9078 | 2.43 | 7.87 | 0.55 | 1.45E–5 | 0.0073 | –0.98 | 7.86 | 1.79E–3 | ... | Single | Spot | Good |
| 2968811 | 727103200009083 | 285.4508 | 38.1172 | 14.85 | 3.94 | 0.26 | 1.99E–5 | 0.039 | –0.93 | 4.05 | 4.1E–4 | ... | Single | Spot | Good |
| 11287726 | 765109200005938 | 286.5219 | 49.0691 | 4.62 | 7.08 | 0.69 | 2.96E–9 | 0.0147 | –0.99 | 7.09 | 1.12E–11 | ... | Single | Spot | Good |
| ... | ... | ... | ... | ... | ... | ... | ... | ... | ... | ... | ... | ... | ... | ... | ... |

[1] https://doi.org/10.5281/zenodo.15336130

This paper has been typeset from a TEX/LATEX file prepared by the author.